\documentclass[preprint]{imsart}

\usepackage[OT1]{fontenc}
\usepackage[utf8]{inputenc}
\usepackage{amsmath,amsthm,amssymb}
\usepackage{graphicx}
\usepackage[pdftex,colorlinks]{hyperref}
\usepackage{natbib}
\bibpunct{(}{)}{,}{a}{}{;}

\usepackage[ruled,vlined]{algorithm2e}
\usepackage{multirow}
\theoremstyle{definition}
\newtheorem{proposition}{Proposition}
\theoremstyle{remark}
\newtheorem*{remark}{Remark}

\newcommand{\set}[1]{\left\{#1\right\}}

\begin{document}

\begin{frontmatter}

  \title{Weighted-Lasso for Structured Network Inference from Time Course Data}

  \runtitle{Structured Dynamic Regulation Network inference}
  
   \begin{aug}
     \author{\fnms{Camille} \snm{Charbonnier}},
     \author{\fnms{Julien} \snm{Chiquet}},
     \author{\fnms{Christophe} \snm{Ambroise}}.
     \ead[label=e1]{christophe.ambroise}
     \ead[label=e2]{camille.charbonnier}
     \ead[label=e3]{julien.chiquet@genopole.cnrs.fr}
     \ead[label=u]{http://stat.genopole.cnrs.fr}
    
     \address{Laboratoire Statistique et G\'enome\\
       523, place des Terrasses de l'Agora\\
      91000 \'Evry, FRANCE\\
       \printead{e3,e2,e1}\\
       \printead{u}} 
    
     \affiliation{CNRS  UMR 8071 \& Universit\'e d'\'Evry}
     \runauthor{Charbonnier, Chiquet, Ambroise}
   \end{aug}
  
  \begin{abstract}
    We present a weighted-\textsc{Lasso} method to infer the
    parameters of a first-order vector auto-regressive model that
    describes time course expression data generated by directed
    gene-to-gene regulation networks.  These networks are assumed to
    own prior internal structures of connectivity which
    drive the inference method. This prior structure can
    be either derived from prior biological knowledge or inferred by
    the method itself. We
    illustrate  the performance of this structure-based penalization
    both on synthetic  data and  on two canonical regulatory networks,
    first yeast cell cycle regulation network by analyzing Spellman
    et al's dataset and second \emph{E. coli} S.O.S. DNA repair
    network by analysing U. Alon's lab data.
  \end{abstract}
    

\end{frontmatter}

\section{Introduction}

Along the dozen of years of statistical studies related to microarrays
for  gene  expression   profiling,  conditional  dependency  has  been
recognized  as  an  appropriate   statistical  tool  to  model  direct
interactions  between  genes.  Graph  representation  suits well  such
relationships  between  variables.  As  a  consequence GGMs  (Gaussian
Graphical   Models)  have  been   widely  studied   by  statisticians,
particularly those  looking for applications to  the reconstruction of
gene-to-gene          regulation          networks          \citep[See
e.g.][]{2005_SAGMB_Schafer,2006_AS_Meinshausen,2006_SAGMB_Wille,2006_JMLR_Castelo,2007_SS_Drton,2007_GI_Shimamura}. In
the  context  of  transcriptomic  data,  the  main  statistical  issue
paradoxically  relies on  the scarcity  of data:  despite  a shrinking
cost,   microarrays  still   provide  dataset   that  fall   into  the
high-dimensional setting.   Namely, the  number of variables  (the $p$
genes)  remains  greater than  the  sample  size  $n$ (the  number  of
microarray slides).

In the Gaussian independent identically distributed (hereafter i.d.d.)
setting, each microarray experiment  is considered as a realization of
a Gaussian  vector whose dependency  structure is fully  determined by
its  covariance matrix.   It can  be shown  that  non-null conditional
dependencies  between genes are  described by  nonzero entries  of the
inverse  of  the  covariance matrix  \citep{1972_Biometrics_Dempster}.
Thus, inferring this  matrix is equivalent to recovering  the graph of
interest, which  is not  trivial when  $n$ is smaller  than or  of the
order  of  $p$.  To  handle  the  data  scarcity, methods  based  upon
$\ell_1$-norm  are very  popular:  they answer  to  both questions  of
regularization  and  of  variable  selection  by  selecting  the  most
significant edges between genes in the network.  In the i.i.d setting,
$\ell_1$-penalized  maximum likelihood Gaussian  covariance estimation
has  been  first   investigated  by  \citet{2007_Biometrika_Yuan}  and
\citet{2008_JMLR_Banerjee}   independently.   These   methods  provide
sparse   graph   estimates,  sparsity   being   a  characteristic   of
gene-to-gene regulation networks.

Looking for  an improvement of these methods  regarding the biological
context, we  provided in  \citet{2009_EJS_Ambroise} a method  that not
only looks for sparse solutions, but also for an internal structure of
the network that drives the inference. Indeed, biological networks and
particularly gene regulation networks are known not only to be sparse,
but  also  organized, so  as  nodes  belong  to different  classes  of
connectivity.   Thus,  we  suggested   a  criterion  that  takes  this
heterogeneity  into account.  This  leads to  a better  inference when
networks  are highly  structured.   Note that  \citet{2009_UAI_Marlin}
published subsequently an independent paper providing a similar method
in a Bayesian framework.  In  these two papers, the internal structure
considered relies  on \emph{affiliation networks}. That  is, genes are
clustered into groups that share the same connectivity patterns.  This
can   be   seen  as   the   analogous   to  the   group-\textsc{Lasso}
\citep{2006_JRSS_Yuan} applied to a graphical context.

Finally,                          some                         authors
\citep[e.g.][]{2007_BMC_Opgen,2009_SAGMB_Lebre,2009_BMCSB_Shimamura}
underlined that transcriptomic dataset are not i.i.d. when considering
time   course   expression  data.   Assuming   a  first-order   vector
auto-regressive (VAR1) model for the time course data generation, they
provided   inference  methods   handling   high-dimensional  settings:
\citeauthor{2007_BMC_Opgen}  suggested   a  shrinkage  estimate  while
\citeauthor{2009_SAGMB_Lebre}    performed   statistical    tests   on
limited-order partial correlations to  select significant edges.  In a
recent work,  \citet{2009_BMCSB_Shimamura} proposed to  deal with this
VAR1  setup by  combining ideas  from  two major  developments of  the
\textsc{Lasso} to define the  Recursive elastic-net. As an elastic-net
\citep{2005_JRSS_Zou},  this method  adds an  $\ell_2$ penalty  to the
original  $\ell_1$ regularization,  thus encouraging  the simultaneous
selection  of highly  correlated covariates  on top  of  the automatic
selection  process  due to  the  $\ell_1$  norm.  As in  the  adaptive
\textsc{Lasso}  \citep{2006_JASA_Zou}, weights  are  corrected on  the
basis of a former estimate so as to adapt the regularization parameter
to  the  relative importance  of  coefficients.   Note  that, in  this
context, we  are no longer looking  for an estimate of  the inverse of
the covariance matrix  but of the parameters of  the VAR1 model, which
leads to a \emph{directed graph}.

In this paper,  we aim to couple the time course  data modeling by the
VAR1  model  to  an  $\ell_1$-regularizing  approach  that  takes  the
internal  structure  of  the  network  into  account.   This  internal
structure  does not  rely on  an affiliation  structure  anymore since
graphs   inferred  from   time  course   data  display   a  completely
asymmetrical pattern.  The internal  structure adopted here splits the
genes  into two groups:  a group  of \emph{hubs}  that exhibit  a high
connection probability to all other genes and a group of \emph{leaves}
that only receive edges leaving  from the hub class.  This information
can  either  be inferred  as  seen in  this  paper  or recovered  from
biological  expertise  since   recovering  hubs  consists  roughly  in
exhibiting  \emph{transcription  factors}  in regulatory  networks,  a
large number of them being already identified by the biologists.
 
Another   refinement   of   our    method   is   to   built   on   the
adaptive-\textsc{Lasso}   \citep{2006_JASA_Zou,2009_A_Zhou}  which  is
known  to  reduce  false  positive  rate  compared  to  the  classical
\textsc{Lasso}.  As such,  our method belongs to the  larger family of
weighted-\textsc{Lasso} methods.  \citet{2007_GI_Shimamura} built upon
\citeauthor{2006_AS_Meinshausen}'s  neighborhood   selection  and  the
adaptive \textsc{Lasso} to improve  inference of networks in an i.i.d.
context.  They  chose separate penalties for  each node's neighborhood
selection problem  and adapted each individual  penalty coefficient to
the information brought by an initial ridge estimate. Here, we suggest
to lower the bias of  the \textsc{Lasso} by not only using information
from an  initial statistical inference  but also from  prior knowledge
about the topology of the  network that assumes the existence of genes
with high connection probability to other genes.

The rest  of the paper is  organized as follows: in  the next Section,
the  VAR1 model  and the  associated likelihood  function  are briefly
recalled;  an  $\ell_1$-penalized  criterion  is proposed  where  each
parameter of  the VAR1 model,  representing the graph of  interest, is
weighted according to its belonging  to the hub group. The weights can
also   depend    on   a   previous    estimate   just   as    in   the
adaptive-\textsc{Lasso}.   We also briefly  recall available  tools to
guide the  choice of the  regularization parameter. In Section  3, the
inference procedure is detailed: we present how the internal structure
can  be recovered;  from that  point, network  inference reduces  to a
convex  optimization  problem which  we  solve  through an  active-set
algorithm  based  upon   the  approach  of  \citet{2000_JCGS_Osborne}.
Finally, an experimental Section  investigates the performances of the
method.  First, simulated data are considered; then, we try to recover
edges implied  in two different regulation processes.   First in yeast
cell  cycle, by  analyzing the  Spellman's dataset  and  comparing the
selected edges to the  direct regulations collected from the Yeastract
database; second  in \emph{E.  coli}, by analyzing  U.  Alon's precise
kinetic data on S.O.S.  DNA repair subnetwork.
\begin{remark}
  The code will soon be embedded in the \texttt{R} package \texttt{SIMoNe}.
  \citep{2008_BI_Chiquet}.
\end{remark}


\section{Modeling Heterogeneous Regulation Networks from Time Course Data}

\subsection{Auto-regressive Model and Sparse Networks}

Let $\mathcal{P} = \left\{1,2,\dots,p\right\}$ be the set of variables
of   interest,   e.g.,   some   $p$   genes.    Let   us   denote   by
$(X_t)_{t\in\mathbb{N}}$ the  $\mathbb{R}^p$-valued stochastic process
that  represents the  discrete-time evolution  of the  gene expression
levels, written as a row vector. Also denote by $X_t^i$ the expression
level of gene $i$ at  time $t$ and $X_t^{\backslash i}$ the expression
level of all genes but $i$ at time $t$. Herein, $X_t$ is assumed to be
generated by a first-order vector auto-regressive (VAR1) model
\begin{equation*}
  X_t = X_{t-1} \mathbf{A}+ \mathbf{b} + \varepsilon_t, \qquad t\in\mathbb{N}^*,
\end{equation*}
where  $\mathbf{A}=(A_{ij})_{i,j\in\mathcal{P}}$  is  an  $p\times  p$
matrix, $\mathbf{b}$ is a size-$p$ row vector and $\varepsilon_t$ is a
white     Gaussian    process.      Namely,     $\varepsilon_t    \sim
\mathcal{N}(0,\mathbf{D})$  where $\mathbf{D}$  is  a diagonal  matrix
such           as           $\mathbf{D}_{ii}=\sigma_i^2$           and
$\mathrm{cov}(\varepsilon_t,\varepsilon_s)    =   \mathbf{1}_{\{s=t\}}
\mathbf{D}$  for all  $s,t>0$.  Moreover,  $X_0  \sim \mathcal{N}(\mu,
\mathbf{\Sigma})$,  with   $\mu$  a  size-$p$  vector   of  means  and
$\mathbf{\Sigma}$   a    covariance   matrix.    Also    assume   that
$\mathrm{cov}(X_t,\varepsilon_s)=0$  for all  $s>t$:  hence, $X_t$  is
obviously a first-order Markov process.

Since  the  covariance matrix  $\mathbf{D}$  is  diagonal, each  entry
$A_{ij}$  is directly  linked to  the partial  correlation coefficient
between variables $X_t^i$ and $X_{t-1}^j$. In fact,
\begin{equation*}
  A_{ij}           =           \frac{\mathrm{cov}\left(X_t^j,X_{t-1}^i          |
      X_{t-1}^{\backslash         i}        \right)}{\mathrm{var}\left
      (X_{t-1}^i|X_{t-1}^{\backslash i}\right)},
\end{equation*}
thus nonzero entries of $\mathbf{A}$  code for the adjacency matrix of
a directed  graph describing the conditional  dependencies between the
elements  of $\mathcal{P}$.  Inferring  $\mathbf{A}$ is  equivalent to
reconstructing this graph and is the main issue of this paper.

To this end, let us set up the estimation framework: assume that $X_t$
is observed on the  time space $t=0,1,\dots,n$. Denote by $\mathbf{X}$
the  $(n+1)  \times  p$   matrix  of  available  centered,  scaled  to
unit-variance data, whose $t^\mathrm{th}$ row contains the information
relative   to  the  $p$   variables  at   time  $t$.    The  empirical
variance--covariance  matrix $\mathbf{S}$  and the  empirical temporal
covariance matrix $\mathbf{V}$ are then given by
\begin{equation*}
  \mathbf{S}   =   \frac{1}{n}\mathbf{X}_{\backslash n}^\intercal
  \mathbf{X}^{}_{\backslash n},\quad    \mathbf{V}    =    \frac{1}{n}
  \mathbf{X}^\intercal_{\backslash n} \mathbf{X}^{}_{\backslash 0},
\end{equation*}
where $\mathbf{X}_{\backslash k}$ denotes matrix $\mathbf{X}$ deprived
of its $k^\mathrm{th}$ row.

The well-known  maximum likelihood estimator (MLE)  of $\mathbf{A}$ is
easily recovered and recalled in the following proposition.

\begin{proposition}\label{prop:var1_mle} Maximizing the log-likelihood
  of the VAR1 process
  is equivalent to the following maximization problem
  \begin{equation*}
    \max_{\mathbf{A}} \left\{\mathrm{Tr} \left(\mathbf{V}^\intercal \mathbf{A}\right)
      - \frac{1}{2}\mathrm{Tr}\left(\mathbf{A}^\intercal   \mathbf{S}
        \mathbf{A}\right) \right\},
  \end{equation*} 
  whose solution is given by
  \begin{equation}
    \label{eq:mle}
    \widehat{\mathbf{A}}^\mathrm{mle} =  \mathbf{S}^{-1}\mathbf{V} .
  \end{equation}
\end{proposition}

\begin{remark}Thanks to  the assumptions we made  on $\varepsilon$ the
  VAR1  model can be  seen as  a usual  regression problem:  denote by
  $\mathbf{X}_p$   (respectively   $\mathbf{X}_f$)   the   $n$   first
  (respectively        last)        rows       of        $\mathbf{X}$.
  $\hat{\mathbf{A}}^{\text{ols}}$     is     naturally    given     by
  $(\mathbf{X}_p^\intercal     \mathbf{X}_p)^{-1}\mathbf{X}_p^\intercal
  \mathbf{X}_f                                                        =
  \mathbf{S}^{-1}\mathbf{V}=\hat{\mathbf{A}}^{\text{mle}}$.   The  MLE
  \eqref{eq:mle} is straightforwardly equivalent to the ordinary least
  square estimate (OLS) of $\mathbf{A}$.
\end{remark}

Solution \eqref{eq:mle} requires a covariance matrix $\mathbf{S}$ that
is invertible, which occurs when  $\mathbf{S}$ is at least of rank $p$.
In real situations the number  of observations is often about or lower
than  the number  of  variables,  thus MLE  needs  to be  regularized.
Regularization   such    as   Moore-Penrose   pseudo    inversion   or
$\ell_1$-regularization can be applied on matrix $\mathbf{S}$ in order
to  make the  inversion  always achievable.   A  sharpest approach  is
investigated  in  \citet{2007_BMC_Opgen}, where  the  OLS solution  is
regularized by shrinking both matrices $\mathbf{S}$ and $\mathbf{V}$.

We suggest to draw  inspiration from the $\ell_1$-penalized likelihood
approach developed by \citet{2008_JMLR_Banerjee} in the case of i.i.d.
samples of a multivariate  Gaussian distribution: here, samples are no
longer i.i.d  yet linked through time  by the VAR1  model.  Still, the
sparsity can be  controlled with a positive scalar  $\rho$ adjoined to
an $\ell_1$-norm penalty on $\mathbf{A}$ by solving
\begin{equation}
  \label{eq:mle_penalized}
  \hat{\mathbf{A}}^{\ell_1}  = \arg\max_{\mathbf{A}} \left\{
    \mathrm{Tr}       \left(\mathbf{V}^\intercal      \mathbf{A}\right)      -
    \frac{1}{2}\mathrm{Tr}\left(\mathbf{A} ^\intercal  \mathbf{S}
      \mathbf{A}\right) - \rho \|\mathbf{A}\|_{\ell_1}\right\}.
\end{equation}
Since MLE  and OLS are equivalent  in this framework,  solution to the
penalized-likelihood     formulation    \eqref{eq:mle_penalized}    is
equivalent to solving $p$  independent \textsc{Lasso} problems on each
column       of       $\mathbf{A}$,       which       is       exactly
\citeauthor{2006_AS_Meinshausen}'s  approach.  The difference  is that
it does not require any post-symmetrization since there is no symmetry
constraint on $\mathbf{A}$ in the present context.

\subsection{A Structured Modeling of the Network}

To attempt a better fit of  data, we suggest that $\mathbf{A}$ owns an
internal structure that describes  classes of connectivity between the
variables.   Indeed,  the  $\ell_1$-norm regularization  encourages  a
first restriction on the  network's topology inferred through criteria
\eqref{eq:mle_penalized},  by encouraging sparsity.   Yet, it  is well
known that by penalizing  truly significant entries of $\mathbf{A}$ as
much as  truly zero  entries a single  $\ell_1$ penalization  leads to
biased estimates  and a particularly strong number  of false positives
\citep{2000_AoS_Fu,2006_JASA_Zou}.  Weighted-\textsc{Lasso} approaches
can lower this  bias by adapting penalties to  prior information about
where the true zero entries should be, relying on possibly data-driven
as well as biological information.  An existing correction is given by
the Adaptive-\textsc{Lasso} \citep{2006_JASA_Zou,2009_A_Zhou}. Penalty
coefficients  are  alleviated or  increased  using individual  weights
reversely      proportional      to      an      initial      estimate
$\mathbf{A}^{\text{init}}$.

The main purpose of this paper  is to show the interest of taking into
account information about the topology of the network: not only should
we scale  coefficients individually, but also  consider the underlying
organization  of  $\mathcal{P}$.  Adaptation  of  weights  is made  by
providing $\mathbf{A}$ with  a well-chosen prior distribution, relying
on the organization of $\mathcal{P}$.  We assume that genes are spread
through  a partition  of $\mathcal{P}$  into $\mathcal{Q}$  classes of
connectivity.  Both existences and  weights of edges, described by the
elements of $\mathbf{A}$, depend on the connectivity class each vertex
belongs to.  Denote  by $Z_{iq}$ the indicator function  that gene $i$
belongs  to  class  $q$.   Each  entry  $A_{ij};Z_{iq}Z_{j\ell}=1$  is
provided  with   an  independent  prior   distribution  $f_{ijq\ell}$.
Following  \citet{2009_EJS_Ambroise}, we choose  Laplace distributions
for $f_{ijq\ell}$ since it is the corresponding log-prior distribution
to the $\ell_1$ term in the \textsc{Lasso}. Hence, by choosing
\begin{equation*}
  f_{ijq\ell}(x)               =              \frac{1}{2\lambda_{ijq\ell}}
  \exp\left\{-\frac{|x|}{\lambda_{ijq\ell}}\right\}, 
\end{equation*}
where $\lambda_{ijq\ell}$  are scaling  parameters, we expect  a model
whose    log-likelihood    will     naturally    make    a    specific
$\ell_1$-penalization term appear.

\paragraph*{Modeling hubs.}  Many configurations fit into this general
model.   In  \citet{2009_EJS_Ambroise} we  focused  on an  affiliation
model.   This structure  opposes intra  to  inter-cluster connections,
assuming the  former to be  far more likely  than the latter.   In the
present context, where dynamic  regulatory networks are represented by
directed graphs,  the affiliation model  unnaturally assumes symmetric
probabilities for  ``incoming'' and  ``outgoing'' edges and  should be
banished.   Indeed,  adjacency matrices  associated  to directed  gene
regulatory networks  are asymmetrical: genes belong  to two completely
different groups.   While a group  of hubs exhibits a  high connection
probability  to all  other  genes,  the remaining  set  of genes  only
receives edges  leaving from the  first class. Illustration of  this phenomenon  by
\citet{1998_MBC_Spellman}'s dataset on \emph{Saccharomyces cerevisiae}
is  presented in  Section  \ref{sect:experiments}. This  setup can  be
summarized as follows:
\begin{equation*}
  f_{ijq\ell} = \left\{
    \begin{array}{ll}
      f_{\text{hub}}\left(\cdot;\lambda_{\text{hub}}\right)    &
      \textrm{if $q$ is the hub class,} \\[1ex]
      f_{\text{leaf}}\left(\cdot;\lambda_{\text{leaf}}\right)   &
      \textrm{if $q$ is not the hub class}. 
    \end{array} \right.
\end{equation*}
Note this  structure only  differentiate edges on  the basis  of their
origin,  whether they  leave  from a  hub  or not,  whatever be  their
arrival  points.  In  this type  of structure  built around  hubs, the
number of classes is fixed at 2.

Allowing  for individual  prior information  about $i$  and  $j$, this
model can be generalized to
\begin{equation*}
  f_{ijq\ell} = \left\{
    \begin{array}{ll}
      f_{\text{hub}}\left(\cdot;\lambda_{ij}\lambda_{\text{hub}}\right)    &
      \textrm{if $q$ is the hub class,} \\[1ex]
      f_{\text{leaf}}\left(\cdot;\lambda_{ij}\lambda_{\text{leaf}}\right)   &
      \textrm{if $q$ is not the hub class.}
    \end{array} \right.
\end{equation*}

\paragraph{The likelihood.}  As the matrix $\mathbf{A}$ has been given
a prior distribution, our aim is to maximize the posterior probability
of $\mathbf{A}$,  given the data  $\mathbf{X}$. For a  fixed structure
$\mathbf{Z}$, this is equivalent to maximizing the joint probability
\begin{equation*}
  \hat{\mathbf{A}}=         \arg\max_{\mathbf{A}}        \        \log
  \mathbb{P}(\mathbf{X}, \mathbf{A} ; \mathbf{Z}).
\end{equation*}
Now,  the likelihood $\mathbb{P}(\mathbf{X},  \mathbf{A}; \mathbf{Z})$
is straightforwardly given by
\begin{equation}
  \label{eq:mle_penalized_structure}
  \log \mathbb{P}(\mathbf{X},\mathbf{A};\mathbf{Z}) = \mathrm{Tr}
  \left(\mathbf{V}^\intercal                \mathbf{A}\right)                -
  \frac{1}{2}\mathrm{Tr}\left(\mathbf{A} ^\intercal       \mathbf{S}
    \mathbf{A}   \right)    -   \|   \mathbf{P}^{\mathbf{Z}}   \star
  \mathbf{A}\|_{\ell_1} + c,
\end{equation}
where $c$  is a constant  term and the  $p\times p$ penalty  matrix is
defined by
\begin{equation*}
  \mathbf{P}^{\mathbf{Z}}=(P^{\mathbf{Z}}_{ij})_{i,j\in\mathcal{P}} =
  \sum_{q,\ell\in\mathcal{Q}}  \frac{Z_{iq}Z_{j\ell}}{\lambda_{ijq\ell}}. 
\end{equation*}
Practically, we obtain the following penalty
\begin{equation*}
  \mathbf{P}_{ij}^{\mathbf{Z}}= 
  \lambda^{-1}_{ij}\cdot\left( \lambda^{-1}_{\text{hub}} Z_{i,\text{hub}} + \lambda^{-1}_{\text{leaf}}
    Z_{i,\text{leaf}}  \right) = 
  \rho \cdot \rho_{ij}\cdot\left( \rho_{\text{hub}} Z_{i,\text{hub}} + \rho_{\text{leaf}}
    Z_{i,\text{leaf}}  \right), 
\end{equation*}
where $\rho>0$  is a common factor  to $\lambda^{-1}_{\text{hub}}$ and
$\lambda^{-1}_{\text{leaf}}$,  which  can  vary  so as  to  adapt  the
penalty                 while                 the                ratio
$\lambda^{-1}_{\text{hub}}/\lambda^{-1}_{\text{leaf}}
=\rho_{\text{hub}}/\rho_{\text{leaf}}>1$ remains  constant at a chosen
level.   Coefficient  $\rho_{ij}$ can  be  held  fixed  at 1  when  no
individual  information  is taken  into  account  or  replaced by  any
well-chosen transformation  of an initial estimate  of $\mathbf{A}$ in
order to provide accurate information on where true zeros might be.

\subsection{Tuning the penalty parameter}

We  briefly recall  here  the different  techniques  available in  the
literature. Asymptotic theory  of the \textsc{Lasso} demonstrates that
a  penalty parameter  of order  $\sqrt{n}$ guarantees  both estimation
consistency and  selection consistency: asymptotically,  estimation is
unbiaised and all  relevant covariates are included in  the model with
strictly positive  probability \citep{2000_AoS_Fu}. In  practice, this
does not tell us which penalty to  use for a fixed sample size $n$. To
solve this  problem, \citet{1996_JRSS_Tibshirani} suggests  the use of
cross-validation. However  it is well-known that the  best penalty for
prediction  is not  the  best penalty  for  model selection  purposes.
Cross-validation is therefore unrevelant here.  Optimality in terms of
selection naturally  draws attention towards penalties  which would in
some way  control the false discovery  rate. Closest to  that goal are
penalty  choices which  guarantee a  control over  the  probability of
connecting two nodes by a chain of edges though no such path exists in
the   true   graph.    Such   penalties   have   been   discussed   in
\citet{2008_JMLR_Banerjee},       \citet{2006_AS_Meinshausen}       or
\citet{2009_EJS_Ambroise} for instance.  However, as underlined in the
latter, this kind of penalty is often much too conservative to be used
as  anything  else than  an  upper bound  on  the  set of  interesting
penalties.   Relying   on   the   Bayesian   interpretation   of   the
\textsc{Lasso}, another option is to maximize the marginal probability
of  the  data  over   all  possible  tuning  parameters.   A  specific
approximation  for  graphs  is derived  in  \citet{2007_GI_Shimamura}.
Taking into account the fact that  the number of degrees of freedom of
the  \textsc{Lasso}  equals the  final  number  of nonzero  parameters
\citep{2007_AS_Zou} computations get  a lot easier.  Particularly, the
BIC  approximation of  the marginal  distribution as  well as  the AIC
criterion, whose  good properties for model  selection are well-known,
are  trivial  to  compute.   In  our case,  we  obtain  the  following
expressions:
\begin{align*}
\textrm{BIC}_\rho &= n \left[\mathrm{Tr}
  \left(\mathbf{V}^\intercal                \widehat{\mathbf{A}}_\rho\right)                -
  \frac{1}{2}\mathrm{Tr}\left(\widehat{\mathbf{A}}_\rho ^\intercal       \mathbf{S}
    \widehat{\mathbf{A}}_\rho\right) \right] - \frac{\log(n)}{2} \textrm{df}_\rho, \notag\\
\textrm{AIC}_\rho &= n \left[\mathrm{Tr}
  \left(\mathbf{V}^\intercal                \widehat{\mathbf{A}}_\rho\right)                -
  \frac{1}{2}\mathrm{Tr}\left(\widehat{\mathbf{A}}_\rho ^\intercal       \mathbf{S}
    \hat{\mathbf{A}}_\rho\right) \right] - \textrm{df}_\rho \notag,
\end{align*}
where $\widehat{\mathbf{A}}_\rho$ denotes the estimate of $\mathbf{A}$
associated to penalty $\rho$ and $\textrm{df}_\rho$ the number of
nonzero entries in $\widehat{\mathbf{A}}_\rho$.

In  practice,  we  observe  that  these  last  criteria  present  both
advantages  of  being  straightforward  to compute  and  of  providing
impressively sensible  results in terms  of both Recall  and Precision
rates. We therefore adopt these  criteria to select two best penalties
to choose from in the remaining of the paper.


\section{Inference Strategy}

\subsection{Structure Inference}
 In many application fields, the structure can be considered as known,
 learned from  expert knowledge.  In genetic for  instance, biologists
 can often extract the list  of transcription factors from the overall
 set of target genes. 

 Otherwise,  the structure,  or part  of it,  could remain  latent: we
 suggest  a   basic  strategy  that  performs   well  practically  for
 biological networks.  In this context, the structure goes down to the
 identification of hubs.  To this  purpose we suggest a very intuitive
 path.  A  first matrix $\mathbf{A}_0$ is estimated  using an adequate
 single \textsc{Lasso}  penalty. We  rely on AIC  and BIC  criteria to
 identify the  best initial penalty.   Nodes are then  classified into
 two  groups,  hubs  and  leaves,  according  to  the  values  of  the
 $\ell_1$-norms of the corresponding rows in $\mathbf{A}_0$.  In order
 to account for the  particularly strong heterogeneity between the two
 groups  (differences  in size  and  dispersion),  a Gaussian  mixture
 approach  is  used  for  clustering  the  genes.   This  defines  two
 submatrices   $\mathbf{A}_0^1$    and   $\mathbf{A}_0^2$   containing
 respectively the lines corresponding  to the first and second groups.
 Hubs  are then  characterized  as  the class  with  the maximum  mean
 absolute value of $\mathbf{A}_0^k$.

\subsection{Active-Set algorithm for Network Inference}

Once  the  internal  structure has  been  recovered,
inference  of   $\mathbf{A}$  amounts  to   optimizing  the  penalized
likelihood  \eqref{eq:mle_penalized_structure} where  $\mathbf{Z}$ are
fixed  parameters.    This  can  be  achieved  by   solving  some  $p$
independent \textsc{Lasso}--style problems  since there is no symmetry
constraint   on   $\mathbf{A}$:   denoting   by   $\mathbf{M}^k$   the
$k^{\mathrm{th}}$ column  of a given  matrix $\mathbf{M}$, we  wish to
solve  for  each column  of  $\mathbf{A}$  the following  minimization
problem
\begin{equation}
  \label{eq:mle_A_columnwise}
  \hat{\mathbf{A}}^k  =  \arg \min_{\beta}  L(\beta),  \text{ where  }
  L(\beta)= \frac{1}{2}\beta^{\intercal} \mathbf{S}
  \beta - \beta^{\intercal} \mathbf{V}^k + \left\|
    \Lambda \star \beta\right\|_{\ell_1},
\end{equation}
where $\Lambda = (\mathbf{P}^{\mathbf{Z}})^k$ for clarity purpose.

Solving penalized problem  \eqref{eq:mle_A_columnwise} can be achieved
through various algorithms.  The elegant active-set approach suggested
in \citet{2000_JCGS_Osborne} takes advantage of the sparsity of $\beta$
to   solve   the  equivalent   constrained   problem:  starting   from
$\mathbf{0}_p$  as  an initial  guess,  the  set  of active  variables
$\mathcal{A} = \set{i : \beta_i  \neq 0}$ is updated at various stages
of the algorithm  so as we solve linear systems  with limited sizes to
determine    the    current    nonzero   coefficients    denoted    by
$\beta_{\mathcal{A}}$ herein.  The algorithm stops once the optimality
conditions  derived from  the classical  Karush-Kuhn-Tucker conditions
are satisfied.  In the next  paragraph, we detail an adaptation to the
present  context   of  the  \citeauthor{2000_JCGS_Osborne},  initially
developed for the \textsc{Lasso} for linear regression.
\\

The objective  function $L$ in  \eqref{eq:mle_A_columnwise} is convex,
yet  not  differentiable everywhere  due  to  the $\ell_1$-norm:  from
convex  analysis, $\beta$  is solution  to \eqref{eq:mle_A_columnwise}
iif   $\mathbf{0}_p$   belongs   to   the  subdifferential   of   $L$,
which mainly  forms the optimality  conditions of the
  problem.  Here, the subdifferential is given by
\begin{equation*}
\partial_{\beta}L(\beta) =  \mathbf{S} \beta -  \mathbf{V}^k + \Lambda
\star {\boldsymbol\theta},
\end{equation*}
where  ${\boldsymbol\theta}\in\mathrm{sign}(\beta)$, that  is,
$\theta_i = \mathrm{sign}\left(\beta_i\right)$ if $i \in \mathcal{A}$,
and       $\theta_i       \in       [-1,1]$      if       $i       \in
\bar{\mathcal{A}}$. 

Starting from $\beta=\mathbf{0}_p$, we  select the component $\ell$ of
$\beta$ whose  subgradient absolute value  is maximal: as a  matter of
fact, a subgradient highly  different from zero induces high violation
of the  optimality conditions.  Such  a choice will guarantee  a large
reduction  of  the  objective  function $L$  during  the  optimization
procedure.   Thus,   this  component  is  added  to   the  active  set
$\mathcal{A} = \mathcal{A}\cup\set{\ell}$.

Then,   optimization  is  only   performed  on   nonzero  coefficients
$\beta_{\mathcal{A}}$ whose cardinal is small since the solution is likely
to be  sparse.  This  is done by  minimizing $L(\beta_{\mathcal{A}})$,
which  reduces  to  a   classical  optimization  problem  because  the
subdifferential turns  to an usual gradient $\nabla_{\beta}  L$ on the
active set $\mathcal{A}$.

While   optimizing,   the   next   update   $\beta^+_{\mathcal{A}}   =
\beta_{\mathcal{A}}   +    \mathbf{h}$   is   obtained    by   solving
$\nabla_{\mathbf{h}}              L(\beta_{\mathcal{A}}              +
\mathbf{h})=\mathbf{0}_{|\mathcal{A}|}$, which  leads to the following
descent direction
\begin{equation*}
  \mathbf{h}        =       -\beta_{\mathcal{A}}       +
  \mathbf{S}_{\mathcal{A},\mathcal{A}}^{-1}\left(\mathbf{V}^k_{\mathcal{A}}
    - \Lambda_{\mathcal{A}} \star 
    \mathrm{sign}(\beta_{\mathcal{A}} + \mathbf{h})\right).
\end{equation*}
However  $\mathrm{sign}(\beta_{\mathcal{A}} +  \mathbf{h})$  cannot be
known while computing $\mathbf{h}$ and is consequently approximated by
the    current     sign    of    $\beta_{\mathcal{A}}$     equal    to
${\boldsymbol\theta}_{\mathcal{A}}$:
\begin{equation*}
  \mathbf{h}        \approx       -\beta_{\mathcal{A}}       +
  \mathbf{S}_{\mathcal{A},\mathcal{A}}^{-1}\left(\mathbf{V}^k_{\mathcal{A}}
    - \Lambda_{\mathcal{A}} \star 
    {\boldsymbol\theta}_{\mathcal{A}}\right).
\end{equation*}

Due to  this approximation, we check for  sign-consistency between the
candidate        update       $\beta_\mathcal{A}+\mathbf{h}$       and
${\boldsymbol\theta}_{\mathcal{A}}$.   In case  of  inconsistency, the
descent  direction  is reduced  so  as  $\beta_{\mathcal{A}} +  \gamma
\mathbf{h}$          is          sign         consistent          with
${\boldsymbol\theta}_{\mathcal{A}}$.  This  ends the optimization part
of the algorithm.

Then,  the active set  $\mathcal{A}$ is  updated since  some $\beta_i$
could have been set to zero during the optimization procedure: this is
done  by looking  for vanished  $\beta_i$s,  verifying $\partial_\beta
L(\beta_i)=0$.   Finally,  optimality conditions  are  tested: if  the
maximal $\ell$ of the  subdifferential corresponding to an unactivated
component of $\beta$ is zero, we have found a solution; otherwise, the
active  set is  updated by  adding $\ell$  to $\mathcal{A}$,  since it
induces the highest reduction of $L$.

These  three  steps  ---  optimization,  deactivation  and  optimality
testing  -- are repeated  until a  solution has  been found,  which is
guaranteed  \citep[see][]{2000_JCGS_Osborne}.  The  full  algorithm is
detailed   below.     Note   that    it   can   either    start   from
$\beta^0=\mathbf{0}_p$ or from a solution obtain from a more penalized
problem with larger vector of  penalties $\Lambda$, that speeds up the
computation,  hence   having  a  behavior  that  is   similar  to  the
homotopy/\textsc{Lars} algorithm \citep{2004_AS_Efron}.

\begin{algorithm}\label{algo:active_set}
  \begin{footnotesize}
    \dontprintsemicolon
    
    \CommentSty{//INITIALIZATION}\; 
    $\beta \leftarrow \beta^0, \mathcal{A} \leftarrow \set{i :  \beta_i  \neq 0}, \theta
    \leftarrow \mathrm{sign}(\beta)$\;
    \BlankLine
    
    \While{$\mathbf{0}_p \notin \partial_\beta L(\beta)$}{
      \BlankLine
      \CommentSty{//1. OPTIMIZATION OVER $\mathcal{A}$}\; 
      
      \CommentSty{//1.1 Compute the (approximate) direction $\mathbf{h}$}\;
      \begin{equation*}
        \mathbf{h}   =    -\beta_\mathcal{A}    +    \mathbf{S}_{\mathcal{A},\mathcal{A}}^{-1}
        (\mathbf{V}^k_{\mathcal{A}}-\Lambda_{\mathcal{A}}
        \star \theta_\mathcal{A})
      \end{equation*}
      
      \CommentSty{//1.2 Check for sign consistency}\;
      \eIf{$\mathrm{sign}(\beta_{\mathcal{A}}+\mathbf{h}) \neq \theta_{\mathcal{A}}$}{
        \CommentSty{//Find a solution which is sign-feasible}\;
        $\gamma, k \leftarrow \arg \min_{0<\gamma<1} \set{ \gamma,
          k\in\mathcal{A} : \beta_k + \gamma h_k=0}$\; 
        $\beta_{\mathcal{A}}\leftarrow \beta_{\mathcal{A}}+\gamma\mathbf{h}$\;
      }{
        $\beta_{\mathcal{A}} \leftarrow \beta_{\mathcal{A}} + \mathbf{h}$ \;
      }
      
      \BlankLine
      \CommentSty{//2. LOOK FOR NEWLY ZEROED VARIABLES}\;
      \For{$i         :          \beta_i         =         0$         and
        $\min_{\theta\in\mathrm{\mathrm{sign}(\beta_i)}}|\partial_{\beta_i} L(\beta_i)| = 0$}{
        $\mathcal{A} \leftarrow \mathcal{A}\backslash \set{i}$\;
      }
      
      \BlankLine
      \CommentSty{//3. OPTIMALITY TESTING}\; 
      \CommentSty{// Select $\ell$ providing the highest reduction of $L$}\;
      $\ell  \leftarrow \arg  \max_{i\in\bar{\mathcal{A}}}  \nu_i$, where
      $\nu_i=\min_{\theta\in\mathrm{\mathrm{sign}(\beta_i)}}|\partial_{\beta_i}L(\beta_i)|$\;
      \eIf{$\nu_\ell = 0$}{
        Stop and return $\beta$\;
      }{
        Update the active set: $\mathcal{A} = \mathcal{A} \cup \set{\ell}$\;
      }
    }
    \BlankLine
    \caption{Active-set algorithm}
    \label{algo:constraint}
  \end{footnotesize}
\end{algorithm}

The  full  matrix  $\mathbf{A}$   is  directly  recovered  by  binding
column-wisely the solutions to the $p$ \textsc{Lasso}--style problems.

\begin{remark}
  With  this method,  the  sparsity constraint  only  applies to  each
  column  of $\mathbf{A}$.   This constraint  implies that  if  we use
  $n+1$ time points, $\mathbf{S}$ is of rank $n$ and thus no more than
  $n$ connections  can be activated  by the \textsc{Lasso} at  most in
  each  column (assuming  the  penalty  is low  enough  to accept  the
  activation  of  all  possible  edges).  Consequently,  the  sparsity
  constraint  only  applies  to   \emph{incoming}  edges  and  not  to
  \emph{outgoing} ones.   In that sense,  sparsity assumptions implied
  by $\ell_1$ penalization only assume  that each node is regulated by
  a small  set of nodes  and do not  contradict the existence  of hubs
  regulating a huge set of nodes.
\end{remark}


\section{\label{sect:experiments}Experiments and Discussion}

In  this section we  apply our  algorithm to  both synthetic  and real
data.   Comparison  is   made   first  within   the   family  of   the
weighted-\textsc{Lasso}.    We  observe   the   performances  of   the
\textsc{Lasso} when associated with a single \textsc{Lasso} penalty or
an  adaptive  penalty.   For  the  adaptive-\textsc{Lasso},  a  single
\textsc{Lasso} penalty is used as  initial estimator.  We then try two
different hub penalties:  one relying only on the  known hub structure
and  a  last  one  inferring   the  hub  structure  from  the  initial
\textsc{Lasso} estimator.  We denote these estimators by \emph{Lasso},
\emph{Adaptive},         \emph{KnwCl},         and        \emph{InfCl}
respectively. Corresponding penalties can be summarized as follows:
\begin{align*}
  P_{ij}^{\textrm{Lasso}} &\varpropto 1 \\
  P_{ij}^{\textrm{Adaptive}} & \varpropto
  \left(\frac{1}{\hat{A}^{\textrm{init}}_{ij}}\vee 1 \right) \\
  P_{ij}^{\textrm{KnwCl}}     &\varpropto    \left(    \rho_{\text{hub}}
    Z_{i,\text{hub}} + \rho_{\text{leaf}}
    Z_{i,\text{leaf}}  \right) \\
  P_{ij}^{\textrm{InfCl}}     &\varpropto    \left(    \rho_{\text{hub}}
    \hat{Z}_{i,\text{hub}} + \rho_{\text{leaf}}
    \hat{Z}_{i,\text{leaf}}  \right),
\end{align*}
where $  x \vee y =  \max \{x,y\}$ and $\hat{Z}$  denotes the inferred
classification. In the  remainder of this section, we
  fix  the  ratio  $\rho_{\text{leaf}}/\rho_{\text{hub}}  =  2$,  thus
  penalizing twice as much nodes labeled as leaves as nodes labeled as
  hubs.  Note  also that  we choose to  maintain the  modification of
adaptive  weights  adopted   in  \cite{2009_A_Zhou}  and  prevent  the
alleviation  of  penalty  parameters.   This trick  ensures  that  the
adaptive  \textsc{Lasso} will  select  a subnetwork  from the  network
inferred  by the  initial  \textsc{Lasso} estimate.   No  edge can  be
included if  it was already  excluded by the \textsc{Lasso}.   In this
way,  the  adaptive  \textsc{Lasso}  guarantees a  decrease  in  false
positives.

Apart from   our   family    of   weighted-\textsc{Lasso}
  proposals,  comparison will  be made  with state-of-the  art network
  inference methods  in a VAR1 setting: the \emph{Shrinkage} method
suggested by \citet{2007_BMC_Opgen},  the Recursive Elastic Net method
(\emph{Renet-VAR})  developed by \citet{2009_BMCSB_Shimamura}  and the
method    based   on   dynamic    Bayesian   networks    proposed   by
\citet{2009_SAGMB_Lebre}  and  available   in  \texttt{R}  within  the
\texttt{G1DBN} package.

Here, the interest  of the inference lies in the  recovery of the true
edges,  in other  words of  whether  the entries  of $\mathbf{A}$  are
correctly identified  as nonzero.  Our estimators are  mainly used for
discriminating nonzero  entries from others.  Quantities  such as True
Positives (TP),  False Positives (FP),  True Negatives (TN)  and False
Negatives  (FN)  summarize  the  performances  of  these  classifiers.
Precision TP/(TP  + FP)  is the  ratio of the  number of  true nonzero
elements  to the  total number  of nonzero  elements in  the estimated
matrix $\hat{\mathbf{A}}$.  Recall TP/(TP + FN) denotes the proportion
of nonzero elements in  $\mathbf{A}$ which were correctly recovered as
nonzero in  the estimation.  Fallout FP/(FP+TN) gives  on the contrary
the  proportion of zero  elements in  $\mathbf{A}$ which  were falsely
declared  as nonzero  in the  estimation.  In  statistical  terms, the
Recall (or Hit Rate) would be the empirical equivalent of the power of
our classification method considered as  a test, while the Fallout (or
False Alarm Rate)  would correspond to the first  type $\alpha$ error.
Note that, in  the context of sparse network  inference, the number of
total positives  is small compared  to the number of  total negatives.
Thus, small  variations of FP and  TP will induce  small variations in
Fallout  and large  variations in  Recall.  Hence,  comparison between
Precision and Recall is generally  more relevant than Fallout / Recall
comparison  in  the  present  sparse  context. This  is  why  we  will
generally choose to  omit Fallout rates when we  need to alleviate the
presentation of results.

These rates are easily  obtained for the \textsc{Lasso} based methods
since they automatically produce null coefficients.  By increasing the
penalty parameter we obtain sparser and sparser graphs.  We start from
a large enough penalty to constrain all coefficients of
$\hat{\mathbf{A}}$ to 0 and decrease
the penalty until we include as many variables as allowed by the ratio
$n/p$. We then select the best penalty from this list as the one
maximizing either the BIC or the AIC criterion.

Like the \textsc{Lasso}, \emph{Renet-VAR} directly implements variable
selection   and  penalty   choice  is   included  in   the  algorithm.
Concerning  \emph{G1DBN},  we  follow  the  author's
  advice to tune the parameters  of the test procedure as described in
  the additional material  of \cite{2009_SAGMB_Lebre}.  When applying
the  \emph{Shrinkage} method  developped by  \citet{2007_BMC_Opgen}, a
supplementary step is required  to transform continuous results into a
binary solution.   We follow \citeauthor{2007_BMC_Opgen}'s  advice and
rely on local false discovery rates.  This provides each edge with an
existence  probability  conditional  on  the  corresponding  entry  in
$\hat{\mathbf{A}}$.   We  declare  as  inferred  edge  any  edge  with
posterior probability  exceeding the threshold of 80\%  as the authors
do.

\subsection{Simulated Data}

\paragraph{Simulation  settings.} To  assess the  performances  of our
approach, we apply  the previous model to a very favorable setup,
where existing  models already perform  quite well.  We  then decrease
the ratio  $n/p$ in order  to observe the  response of each  method to
this increasing lack of  information. On top of that, we consider
graphs of different sizes: small graphs of 20 nodes, larger graphs of 100 nodes and a
setup with 800 nodes. For smaller graphs, we
consider three different amounts of observations: 10, 20 and 40. For
medium sized graphs, we also consider the cases $n=p/2$ and $n=p$ but
omit the case $n=2p$ as unrealistic. The setup $p=800,n=20$ is meant to mimic
\citeauthor{1998_MBC_Spellman}'s dataset.

Simulation of the  VAR1 process is based upon  the simulation strategy
used by \citeauthor{2007_BMC_Opgen} in  order to ease the comparisons,
but introduces  a structure based on  hubs in order  to better reflect
the structure we could expect from  a real data set.  A graph is first
simulated,   with   fixed   numbers   of  nodes   and   edges.    Like
\citeauthor{2007_BMC_Opgen}  we simulate  sparse  graphs, with  $K=2p$
edges.  Nodes  are split  into two groups  according to  a multinomial
distribution with probabilities (0.1,0.9),  leading to 10\% of hubs in
average.   Edges are  then  positioned  in the  graph  according to  a
multinomial distribution, with  85\% of edges from hubs  to leafs, and
the remaining  set within hubs. Exception  is made for  the very large
graph, for which we base the number of edges and their distribution on
\citeauthor{1998_MBC_Spellman}'s  data.  The  matrix  $\mathbf{A}$  is
synthesized on the basis of  this graph: we attribute a random partial
correlation value uniformly  distributed on $[-1,-0.2]\cup [0.2,1]$ to
all nonzero coefficients (corresponding to edges in the graph).

From this  matrix, a VAR1  observation is generated, using  a centered
Gaussian  starting value  and  a centered  Gaussian  noise, both  with
variance  $\sigma^2=0.1$. For computing time reasons, this  is
repeated 500  times for the small graphs, 200 times for medium sized
graphs and 100 times for the large graph.  Results  are
averaged over all samples.
\\

To gain a better insight into the difficulty of these synthesized data
set   for  a   \textsc{Lasso}  estimator,   we  checked   whether  the
\emph{irrepresentability                                     condition}
\citep{2006_JMLR_Zhao,2008_AS_Meinshausen} was  validated in all these
very  simple  simulations.  First,  note  that  the graphical  context
requires the irrepresentability condition  to be validated for each of
the $p$ genes at the same  time, which makes it much more difficult to
hold  than in the  simple regression  context where  it is  an already
strong  hypothesis.  In our  context, since  we solve  $p$ independent
\textsc{Lasso} problems,  we can check the validity  of the hypothesis
in   each  of   these  individual   problems.  For   each   gene,  the
irrepresentability  condition is  tested using  the true  sign pattern
extracted  from  the  corresponding   column  of  the  true  adjacency
matrix.  Thus  the sets  of  relevant  and  irrelevant covariates  are
allowed to vary from one problem to another. Simulating 100 samples of
each simulation  setting, we observed  that even in a  favorable setup
with  twice  as many  observations  as  variables  ($p=20$ genes)  the
irrepresentability condition fails for 30\% of genes in average . With
$p=20$  genes and only  $n=10$ observations  this assumption  fails in
average for 51\% of the genes.  In other words, for around half of the
genes we  cannot expect the  \textsc{Lasso} to recover the  exact sign
pattern.   See  Table  \ref{tab:IC}   for  details.   Admittedly,  the
irrepresentability condition is  a really strong assumption, necessary
and sufficient  for exact sign recovery,  that is to say  not only the
exact neighborhoods (no false  positives, no false negatives) but also
the exact  signs of the  correlations. Yet since the  simulated values
are quite well separated between true zeros and true nonzeros we would
have  expected  that  this   hypothesis  would  have  been  much  more
validated.   Information about  the validity  of  the \emph{restricted
  eigen-value  assumptions}  \citep{2009_AS_Bickel}  would be  greatly
appreciated to compensate for  such pessimistic results, but these are
computationally intractable.   Adaptation of \citet{2008_A_Juditsky}'s
results to the present context could be of great benefit.

\begin{table}[!ht]
\centering
\begin{tabular}{c|cc}
$_{n/p}\diagdown^{p}$ & 20 & 100 \\
\hline
2& 0.30 (0.23) & -\\
1 & 0.41 (0.23)& 0.37 (0.15)\\
1/2 & 0.51 (0.18)& 0.42 (0.12)\\
\end{tabular}
\caption{\label{tab:IC} Average proportion of genes for which the
  irrepresentability condition does not hold and standard error in
  each simulation setting.}
\end{table}

\begin{figure}[!ht]
  \centering
\includegraphics[width=\textwidth]{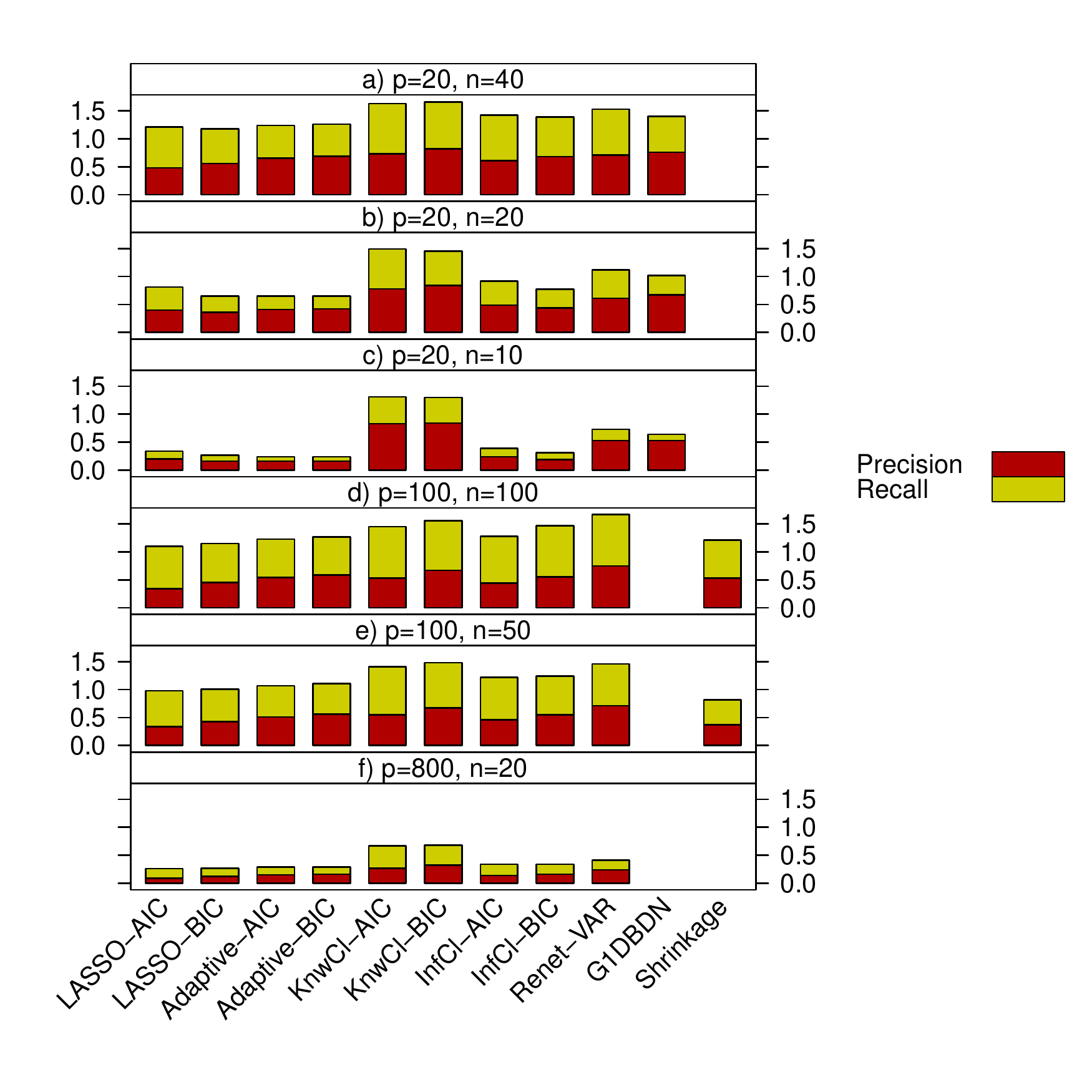}
     \caption{\label{fig:sim} Bar charts of Precision and Recall rates for
        each method and simulation setting, averaged over all simulation
        samples.}
\end{figure}

\begin{figure}[!ht]
\centering
\begin{tabular}{lc}
  \multirow{2}{*}{\rotatebox{90}{$n=100$}} & \includegraphics[width=.7\textwidth]{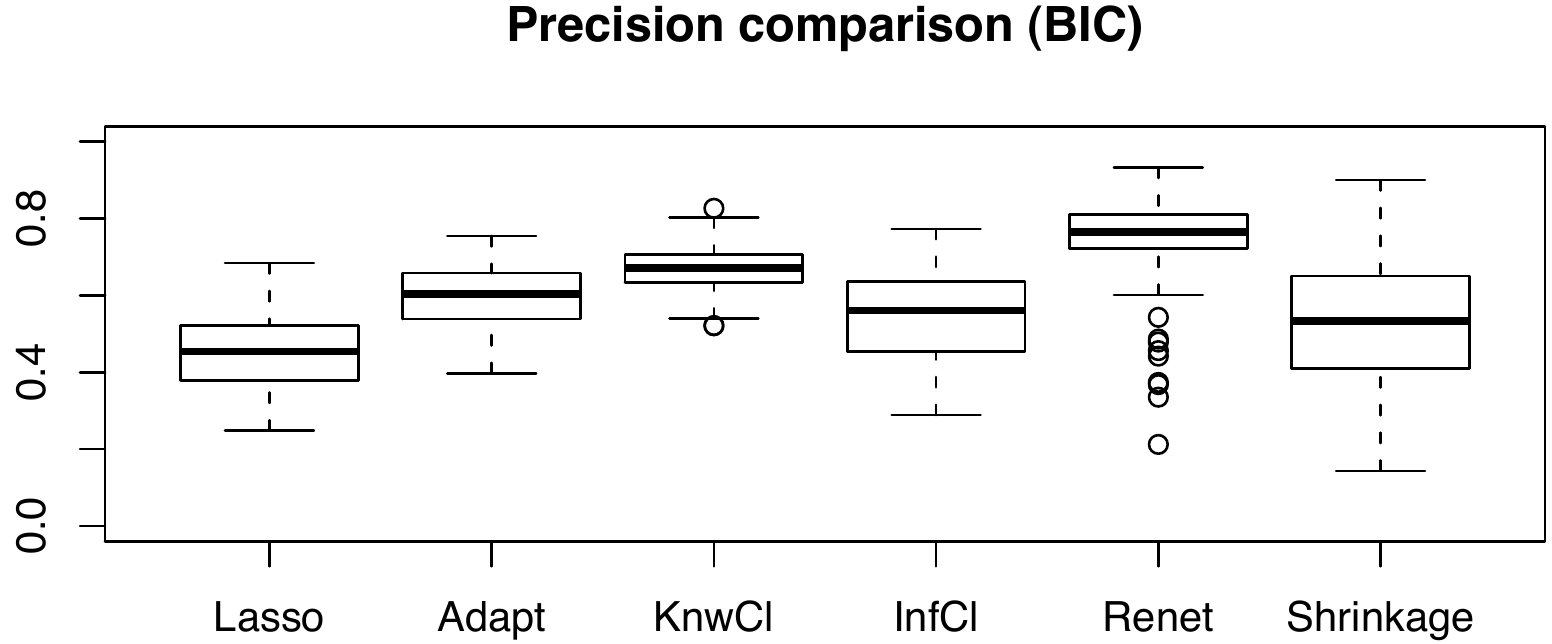}\\
& \includegraphics[width=.7\textwidth]{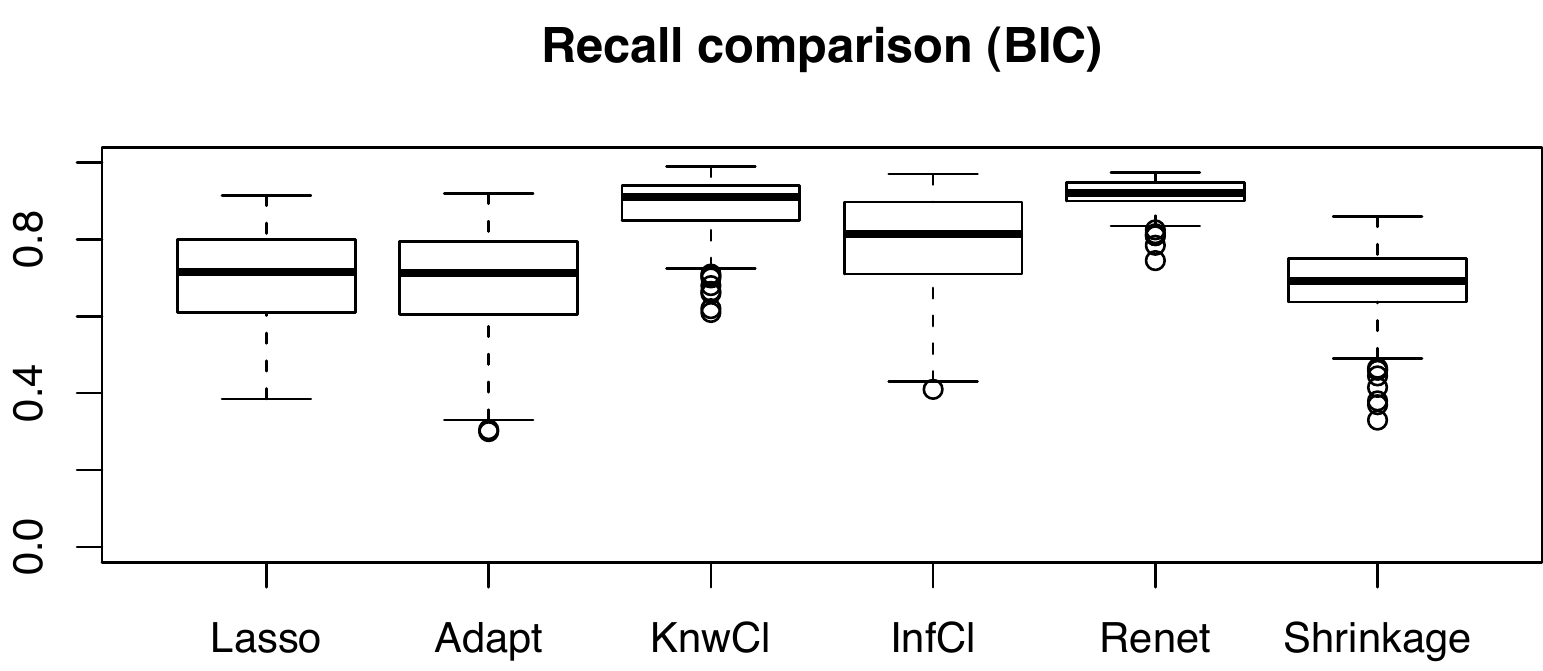}\\[2ex]

\multirow{2}{*}{\rotatebox{90}{$n=50$}} & \includegraphics[width=.7\textwidth]{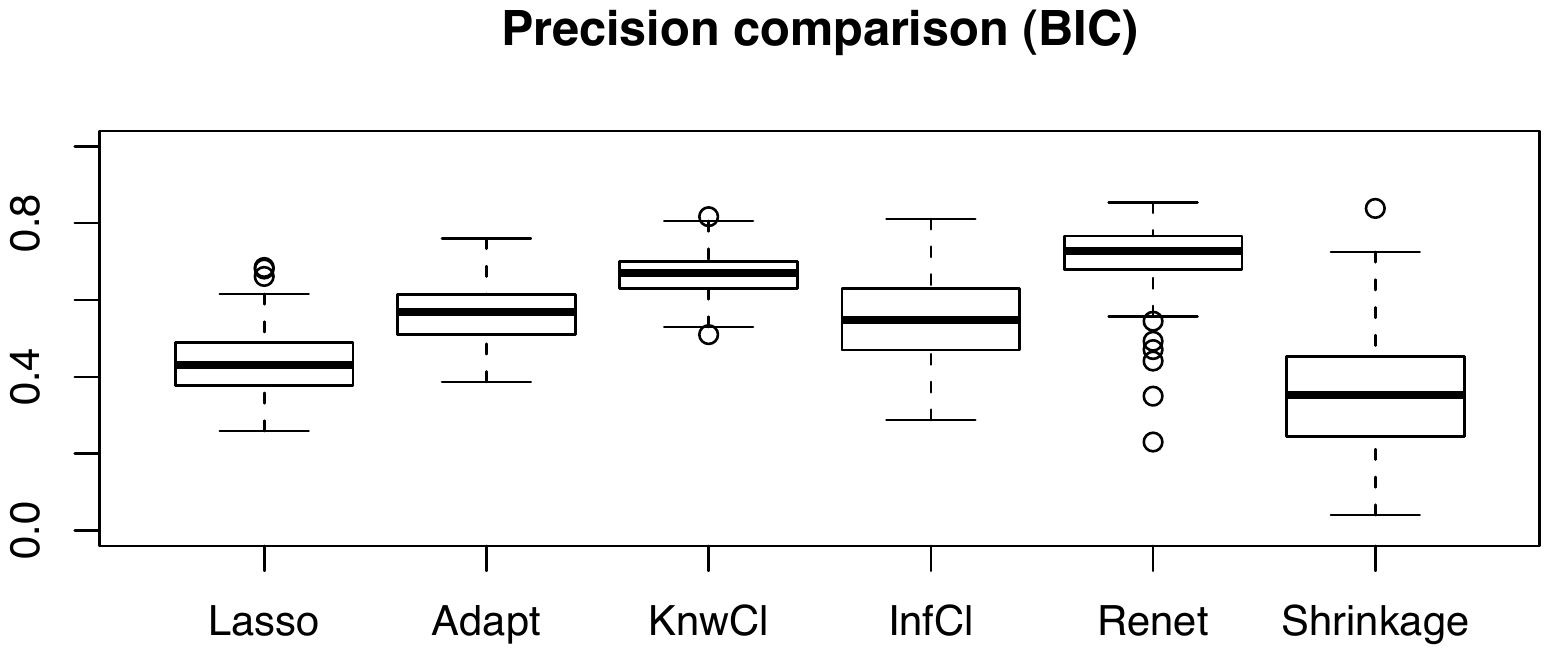}\\
& \includegraphics[width=.7\textwidth]{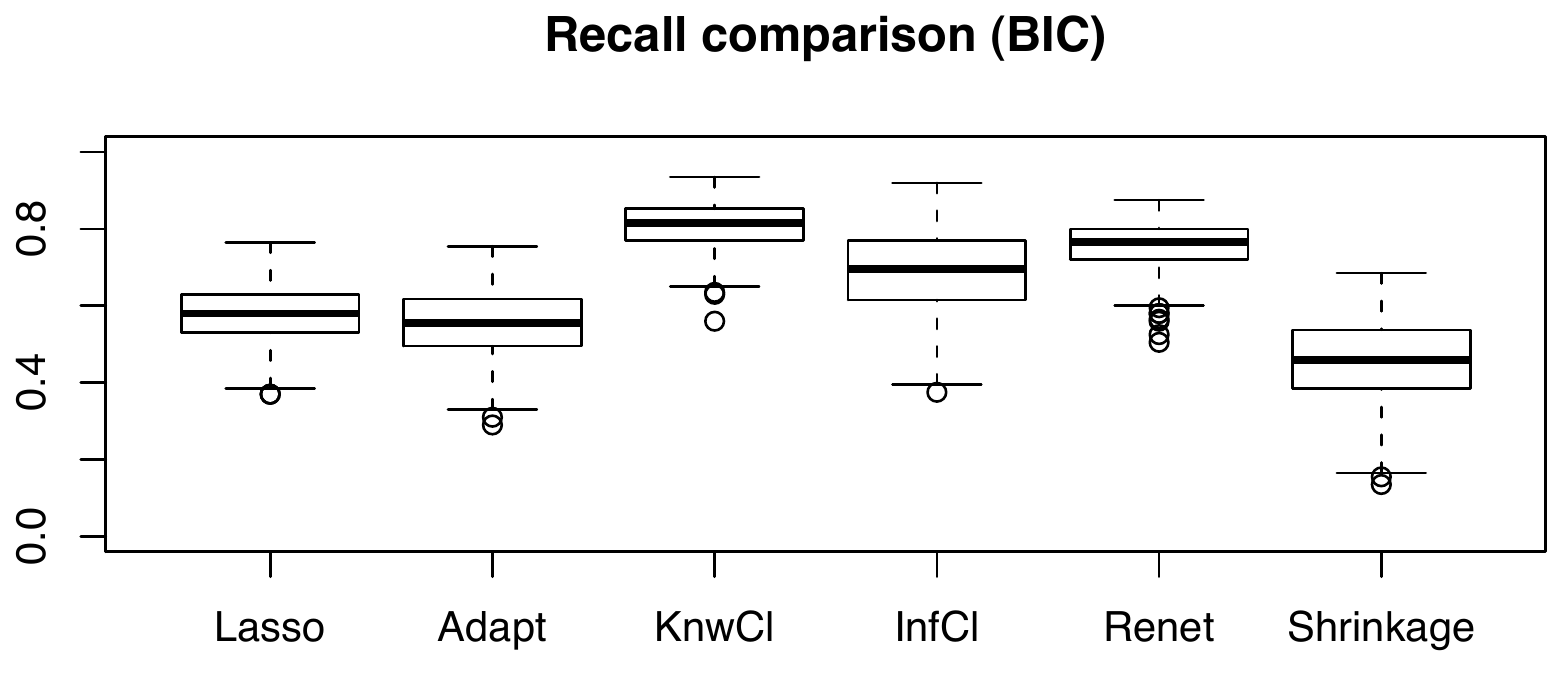}\\

\end{tabular}

\caption{\label{fig:sim100} Boxplots of Precision, Recall and Fallout statistics for all methods except Shrinkage in a setup $p=100$, for 200 simulated data sets. Best Lasso penalties chosen on the basis of the BIC criterion.}
\end{figure}

\paragraph{Discussion of  simulation results.}  Results  are presented
in Figure \ref{fig:sim} under the form of Barcharts.   Figure \ref{fig:sim100} illustrates the case
where $p=100$  by giving boxplots for the  distributions of Precision,
Recall and Fallout.

Compared methods differ with the type of setting. First of all, since the
\emph{Shrinkage} method (particularly the local false discovery rate step) relies on
the hypothesis that $p$ is large, we do not consider it fair to apply
it to the small network setting. Reversely, for computing time reasons
we decided to restrict the application of \emph{G1DBN} to the graphs of size $p=20$.

Penalties  for the  \textsc{Lasso} based  methods were  chosen  on the
basis of either the BIC  or AIC criterion. Although theory states that
the  BIC ought  to  outperform the  AIC  in terms  of model  selection
\citep{2007_AS_Zou}, we  observed that  in practice the  BIC criterion
might be too  conservative when $n$ is small compared  to $p$. In that
situation, it  might be  interesting to favor  the less  stringent AIC
criterion which will induce a higher  recall rate for not such a large
loss in  precision. Note that the  penalty choice based on  the AIC or
the BIC can lead to choose the null model as best model. In that case,
Precision cannot  be defined. We  thus show the results  for precision
over all simulations where at least one variable was included.

The first point worth noting in Figure \ref{fig:sim} is that in all settings the \textsc{Lasso} is outperformed by
weighted-\textsc{Lasso}  methods and others. This   quick  check  confirms  the
interest   of  compensating   for   the  bias   induced  by   $\ell_1$
regularization on  large coefficients.  It is also  possible that what
we  observe about  the  validity of  the irrepresentability  condition
jeopardizes  the performances  of  the single-penalty
\textsc{Lasso}. In line with Table \ref{tab:IC}, the \textsc{Lasso}
performs particularly badly when the ratio $n/p$ is not favorable,
with recall and precision rates under 20\% when $p=20,n=10$. It even
performs so poorly that it deprecates the inference based on
adaptive weights.
\emph{A priori}  information on where  the true zeros might
 compensate for this  apparent lack of ``neighborhood stability'',
using  \citeauthor{2006_AS_Meinshausen}'s vocabulary, and  explain why
the \emph{KnwCl} penalty is far more accurate (precision of
84\% in average for a recall of nearly 50\% in average for the same simulation setting $p=20, n=10$).

As  expected, in all  settings (except  when $n$  is really  too small
compared to $p$) the \emph{Adaptive} penalty improves the precision but at
the  price of a  smaller recall  rate. On  the contrary,  the inferred
classification  \emph{InfCl} allows to  improve the  precision without
undermining the  recall rate.  However,  both methods  are highly
dependent on the initial  \textsc{Lasso} estimate. Therefore, the gain
in  precision resulting  from such  methods decreases  with  the $n/p$
ratio.

Benefitting from a certain amount of supplementary information, the
\emph{KnwCl} penalty leads to a clear increase in both precision and
recall. Particularly when little information is available in terms of
number of observations, taking \emph{a priori} information about which genes are
potential regulators and which are not into account improves the
results dramatically. This is true when compared to all \textsc{Lasso}
based methods but generalizes to \emph{Shrinkage}, \emph{Renet-VAR}
and \emph{G1DBN}. Admittedly, \emph{Renet-VAR} leads to higher
precision values with medium sized graphs, but it is compensated by
smaller recall rates.

Table  \ref{fig:sim} shows naturally  that we  cannot expect  too much
from   very    extreme   settings   ($p=800,n=20$,    that   is,   the
\citeauthor{1998_MBC_Spellman}'s settings). Average Recall rate is less than 20\%
for  all  methods  except  the  \emph{KnwCl} penalty.  In  this  case,
knowledge of potential hubs allows the recall rate to almost double in
average while  increasing the precision.  Note however  that even with
this supplementary information precision rates never exceed 50\%.

\begin{figure}[!ht]
\centering
\includegraphics[width=.6\textwidth]{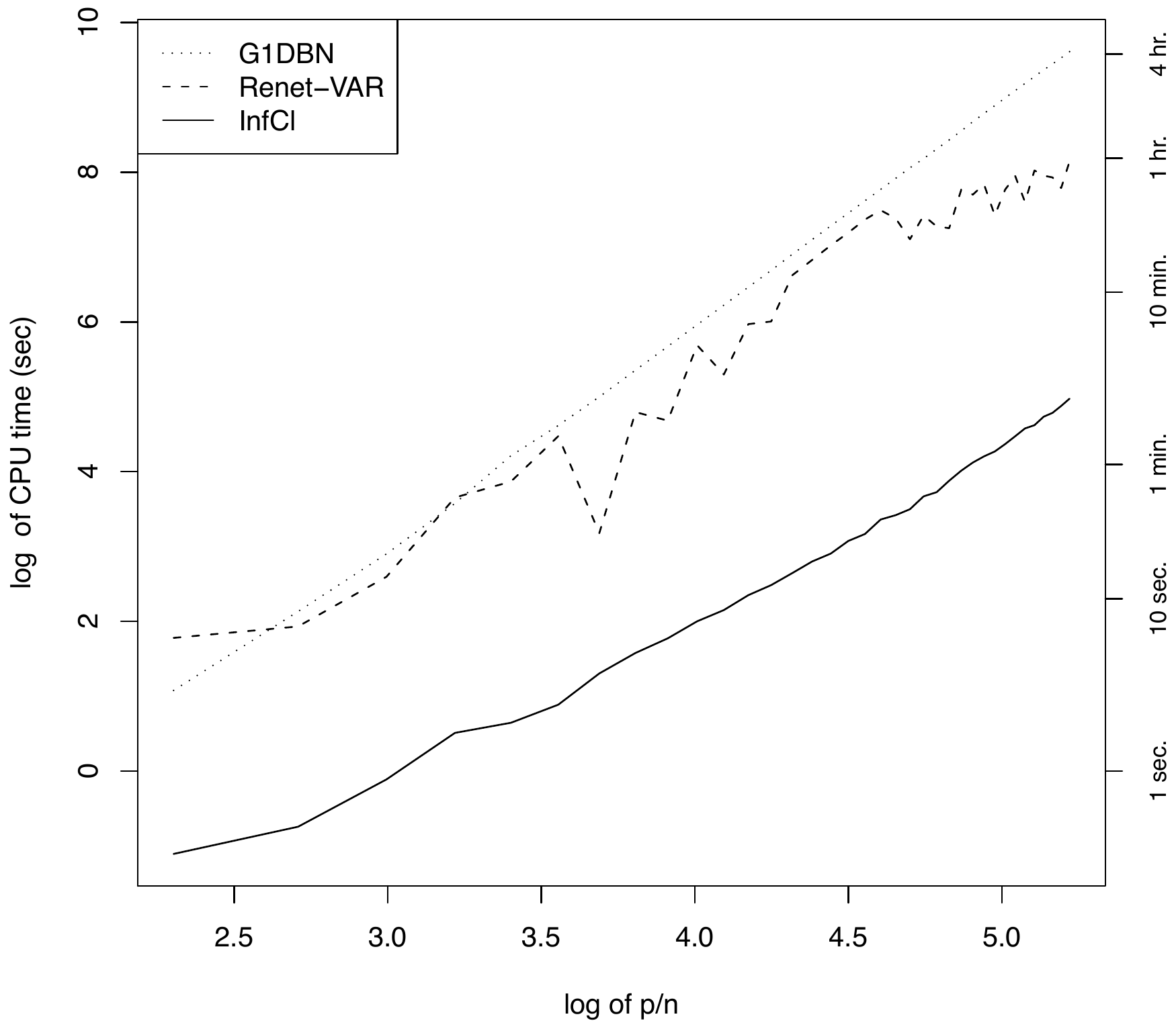}
\caption{\label{fig:cputime} Computing times  on the log-log scale for
  Renet-VAR,    G1DBN    and    InfCl    (including    inference    of
  classes). Intel Dual Core 3.40 GHz processor.}
\end{figure}

To  finish  with, we  would  like to  lay  the  emphasis on  computing
times. For  this we let the  number of nodes  range from 5 to  185 and
fixed the number of observations  at half the maximum number of nodes,
i.e.  $n=92$.  This  leads to  a  ratio  $n/p$  ranging from  0.05  to
2. Computing  times for the weighted \textsc{Lasso}  with inference of
the  classification \emph{InfCl}  and selection  of the  best penalty,
\emph{Renet-VAR} and  \emph{G1DBN} are presented in  the log-log scale
in  Figure  \ref{fig:cputime}.  We  can  see that  running  times  for
\emph{Renet-VAR} and \emph{G1DBN} can become a handicap as soon as $p$
gets  large while  computing times  for \emph{InfCl}  rarely  exceed 2
minutes.

\subsection{Yeast Data}

We confronted  our model  to time measurements  of \emph{Saccharomyces
  cerevisiae}      gene     expression      data      collected     by
\cite{1998_MBC_Spellman}.   We  focus  on  the subset  of  genes  they
identified as  periodic, i.e.   genes whose transcription  levels over
time show evidence that they are cell-cycle regulated.

\paragraph{Remarks on the data set.}  This dataset is one of the first
microarray  experiments.   It  is  thus  doomed to  be  rather  noisy,
contrary  to the  simulated data  sets. Besides,  we had  to  face the
problem  of  missing  values,  which  appeared on  some  of  the  most
important genes. We  imputed them as the mean of  the two closer known
observations in  time for  the gene considered,  before and  after the
time point of interest.

On top  of its noisiness, Spellman  et al.'s data  set is particularly
hard  to tackle  from  a  statistical view point. Information  is
provided  on 786 genes  for only  18 time  points.  This  implies that
using our algorithm we  cannot activate more than $17*786=13362$ edges
out of $789*786=617796$ possible ones, that is to say 2.2\%.

However,  we  can  rely  on  experimental conclusions  on  yeast  gene
regulation networks to collect target information about the true edges
of the graph. We compare  our results to the adjacency matrix provided
by  the  Yeastract   database  (\url{www.yeastract.com}).   We  retain
information on documented direct  relationships, that is to say direct
regulations confirmed by published experimental results.

Note  however  that  this  theoretical  benchmark  is  biased  in  two
ways. First, some true edges  might be missing because all regulations
might not have  been confirmed by experiments yet.  Second, this graph
gathers  all  reported regulations,  whatever  the  conditions of  the
experiment.  Some  might  not   actually  happen  during  the  precise
experiment we consider. We can suppose the effect of the first bias to
be   low   in   a   model   organism   such   as   \emph{Saccharomyces
  cerevisiae}.  The effect  of the  second  bias is  much more  likely
however,  since measurements  are  all  made while  cells  are at  the
beginning of their growth, growing  until ready for DNA synthesis.  We
cannot expect  the whole  range of possible  regulations to  happen in
such a small portion of the cell cycle.

\begin{figure}[htbp!]
  \centering
  \includegraphics[width=.5\textwidth]{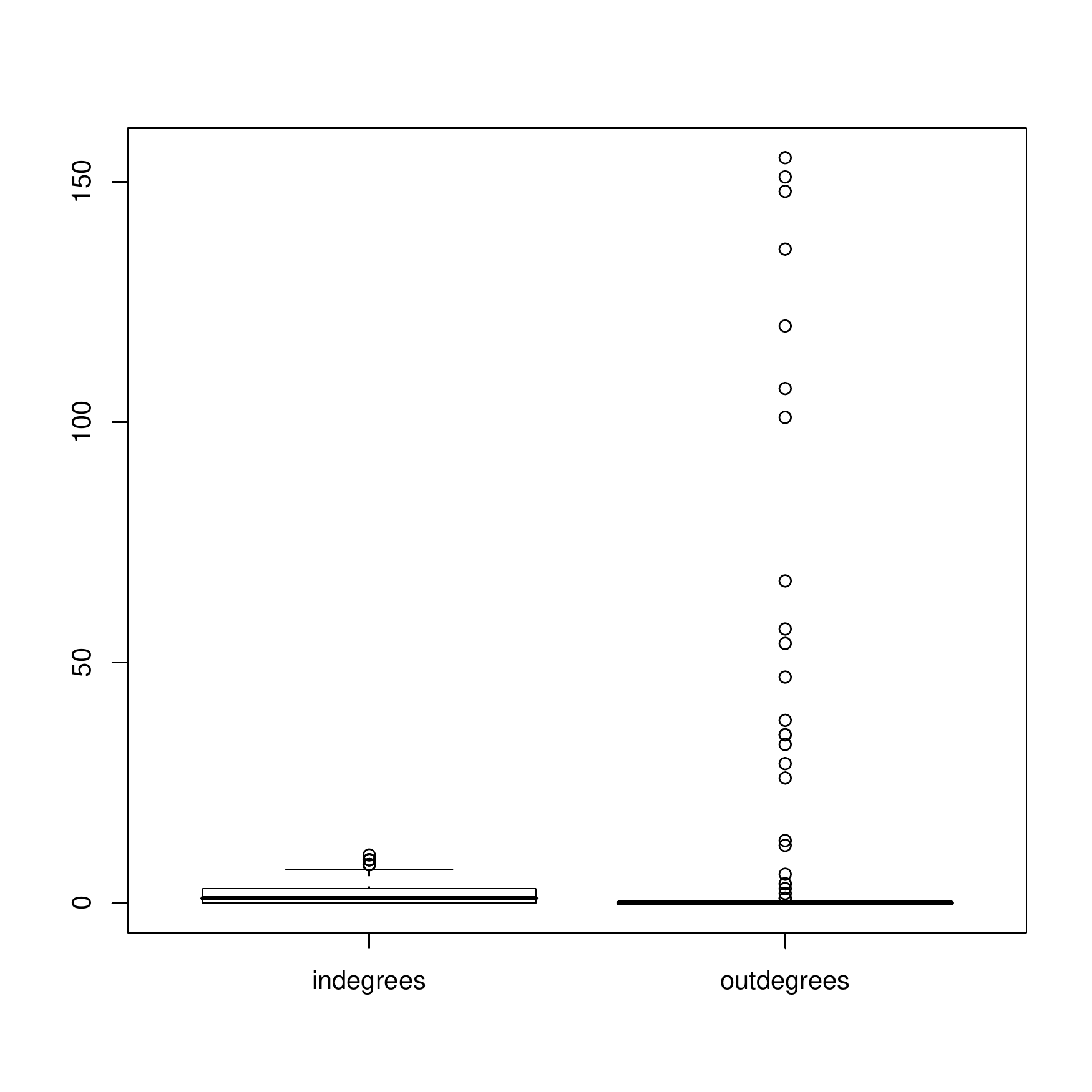}
  \caption{\label{fig:degrees}  Boxplots   of  incoming  and  outgoing
    degrees in Yeast theoretical adjacency matrix }
\end{figure}

This  dataset illustrates  quite  well the  biological properties  our
model  is  based  upon.  First,  documented  information  reveals  the
existence of 1385 true edges  (among more than 600000 possible ones in
theory).  The  theoretical graph is thus  extremely sparse.  Secondly,
the hub structure is quite clear:  edges leave from only 26 out of 786
genes.  Hence  knowledge of the  hubs provides crucial  information on
the position of edges. This  phenomenon also clearly appears on Figure
\ref{fig:degrees}.   Incoming degrees never  exceed 20  but only  1 is
null. On the contrary, outgoing degrees are null for the vast majority
of genes. Significant degrees appear as outliers in this distribution,
reaching up to 150 for some of them.

\paragraph{Discussion of the results.}The  setting is much  harder than in  the first simulated
data sets, with a ratio  $n/p = 2.3\%$ as well as harder than the last
simulated dataset with less separated correlations
between existing  and non existing  edges. 
Results presented in Table \ref{tab:YeastAnalysis} show quite well the
difficulty all methods encounter in front of this data set. Results
for  the \emph{Shrinkage}  approach are  not shown  because  the local
false discovery rate step included in this method was heavily flawed by the lack of
separability between edges and non edges. Except for
the \emph{KnwCl} penalty, all \textsc{Lasso} based estimators are
reduced to the null model. Both the BIC and AIC criteria do not find
the increase in likelihood large enough to compensate for the
complexity of any model with at least one edge. Performances of the
\emph{KnwCl} penalty and \emph{Renet-VAR} remain lower than what we
could expect from simulated results.

\begin{table}[!ht]
  \centering
  \begin{tabular}{lcccccc}
    \hline
    Models & & Lasso & Adaptive & KnownCl & InferCl & Renet \\
    \cline{1-1} \cline{3-7} \\[-1ex]
    Precision & & - & - & 0.082 &  - & 0.004 \\
    Recall & & 0 & 0 & 0.068 & 0 & 0.003 \\
    Fallout & & 0 & 0 & 0.002 & 0 & 0.002 \\
    \hline
\end{tabular}
\caption{\label{tab:YeastAnalysis} Precision, Recall and Fallout performances for all Lasso based methods and Renet-VAR on \citeauthor{1998_MBC_Spellman}'s data set. Best Lasso penalties chosen on the basis of the BIC criterion.}
\end{table}

Many reasons for such bad perfomances could be thought of. We already
mentionned the noisiness of the data, which quite hardly
differentiated the edges from non edges. Second, homogeneity of the
VAR(1) model might be too strong an assumption. Last but not least,
when looking more  closely at how data were  collected we  noticed that measurements  were made  every 7
minutes, which might be long  enough for dependencies to vanish. Also,
since we measure  values related to the cell  cycle, measurements were
necessarily  made on  different cells  each time,  thus  measuring the
expression  levels on  different individuals  at each  time  point. In
brief, this  apparently longitudinal data set might  share more common
points with i.i.d. models than with VAR1 processes.

\subsection{\emph{E. coli} S.O.S. DNA repair network}

In this  section we  quit the high  dimensional setup and  compare the
performances of all methods in a much easier framework.  We focus on a
sub-network from  \emph{E.  Coli} S.O.S.  DNA  repair network analyzed
by  \citeauthor{2002_PNAS_Ronen} \footnote{data  downloadable  on Uri
  Alon's  homepage,  \texttt{http://www.weizmann.ac.il/mcb/UriAlon/}}.
Data provide  information on  the main 8  genes of the  S.O.S. network
(\emph{uvrD,lexA,umuD,recA,uvrA,uvrY,ruvA,polB})    across   50   time
points. Measurements  rely on precise expression  kinetics which allow
\citeauthor{2002_PNAS_Ronen} to monitor  mRNA expression levels every
6 min after exposition  of the DNA to UV light at  time 0. We will not
dwell on  the measurement technology  here (see \cite{2002_PNAS_Ronen}
for details).  Note however that the authors do not measure the actual
mRNA quantity present in the cell at time $t$ but the instant promoter
activity of  each gene.  Equivalence  between the two  measurements is
guaranteed if the instant quantity  of mRNA in the cell roughly equals
its production  rate, that is  to say if  there is no  accumulation of
mRNA in the cell. Under this assumption, \citeauthor{2002_PNAS_Ronen}
's data can be used as any microarray dataset. 

 \emph{E.   coli}  S.O.S.  DNA   repair  network  provides  a  precise
 benchmark: specific regulatory interactions in response to DNA damage
 have been characterized. In other words, we can rely on a theoretical
 regulatory  network which  represents the  main  direct transcriptory
 regulations actually
 taking place during the experiment. According to the regularly
 updated EcoCyc database,
 \emph{lexA} is the only regulator in this subnetwork, regulating all
 genes including itself. Concretly, the protein \emph{LexA} is at the core of
 the  regulation network, usually binding sites  in the  promoter  regions of
 S.O.S.  genes to  repress  their expression.  As  soon as  \emph{RecA}
 senses  DNA damage (by  binding to  single-stranded DNA),  it becomes
 activated and induces  \emph{LexA} autocleavage. The decrease in
 \emph{LexA}   concentration  alleviates   the  repression   of  
 S.O.S.  genes.  When damage is repaired, the level of activated
 \emph{RecA}  drops, \emph{LexA} accumulates  and represses  again all
  S.O.S. genes. 

\begin{figure}[!ht]
\centering
    \includegraphics[width=\textwidth]{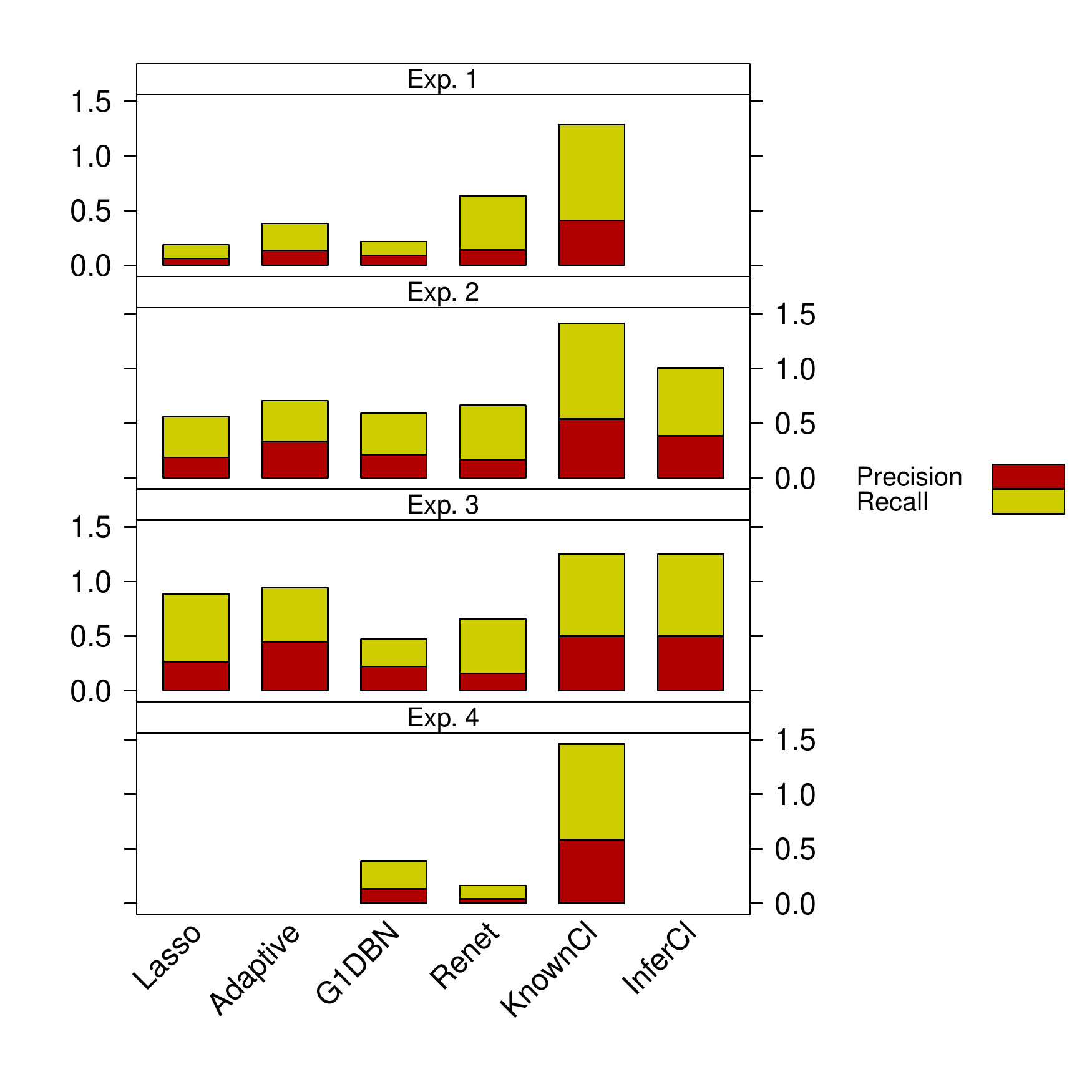}
 \caption{\label{fig:Ecoli} Summary of Precision, Recall and Fallout (respectively Prec., Rec. and Fall.) values for each method and experiment}
\end{figure}

Detailed results are presented in Figure \ref{fig:Ecoli}. We can see
that performances differ a lot from one experiment to another. Particularly,
experiments 1 and 4 lead to significantly poor results although nothing
should \emph{a priori} distinguish them from 2 or 3 (1 and 2, respectively 3 and 4,
share the same U.V. exposure).

As on simulated data, the \textsc{Lasso} leads to poor
results. \emph{G1DBN} shows similarly poor performances here. Quite
surprisingly, \emph{Renet-VAR} does not perform as well as we could have
expected from simulations. It reaches 50\% of recall at the expense of
very low precision rates. \emph{Adaptive}  penalty improves more the quality of the estimation
than in the simulation studies. Now they increase the precision of the
\textsc{Lasso} without really undermining the recall rate. Inference
of the classification outperforms these, with higher recall and
precision rates. This is quite interesting since except in experiments
1 and 4 where the \textsc{Lasso} provide almost no information,
inference of the classes seems quite good although the initial
\textsc{Lasso} still shows mediocre results. To finish with, the
\emph{KnwCl} penalty benefits quite well here from its extra
information since it outperforms all other methods and manages to
reach honest results even in datasets 1 and 4 which disturbed all other methods.

\begin{figure}[!ht]
\centering
\begin{small}
  \begin{tabular}{c@{\hspace{.5cm}}c}
    \textsf{Lasso} & \textsf{Adaptive Lasso} \\
    \includegraphics[width=.4\textwidth]{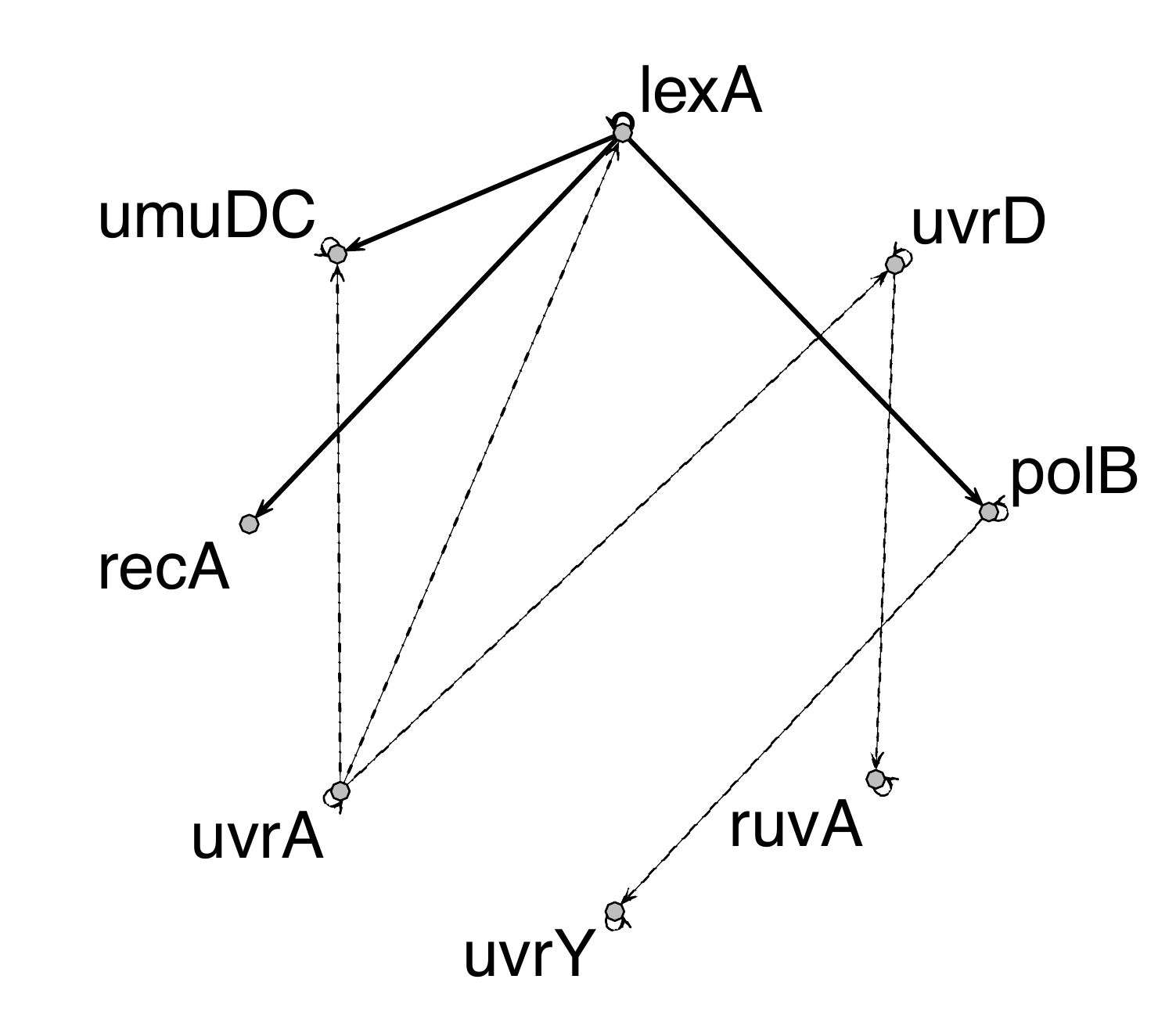} & \includegraphics[width=.4\textwidth]{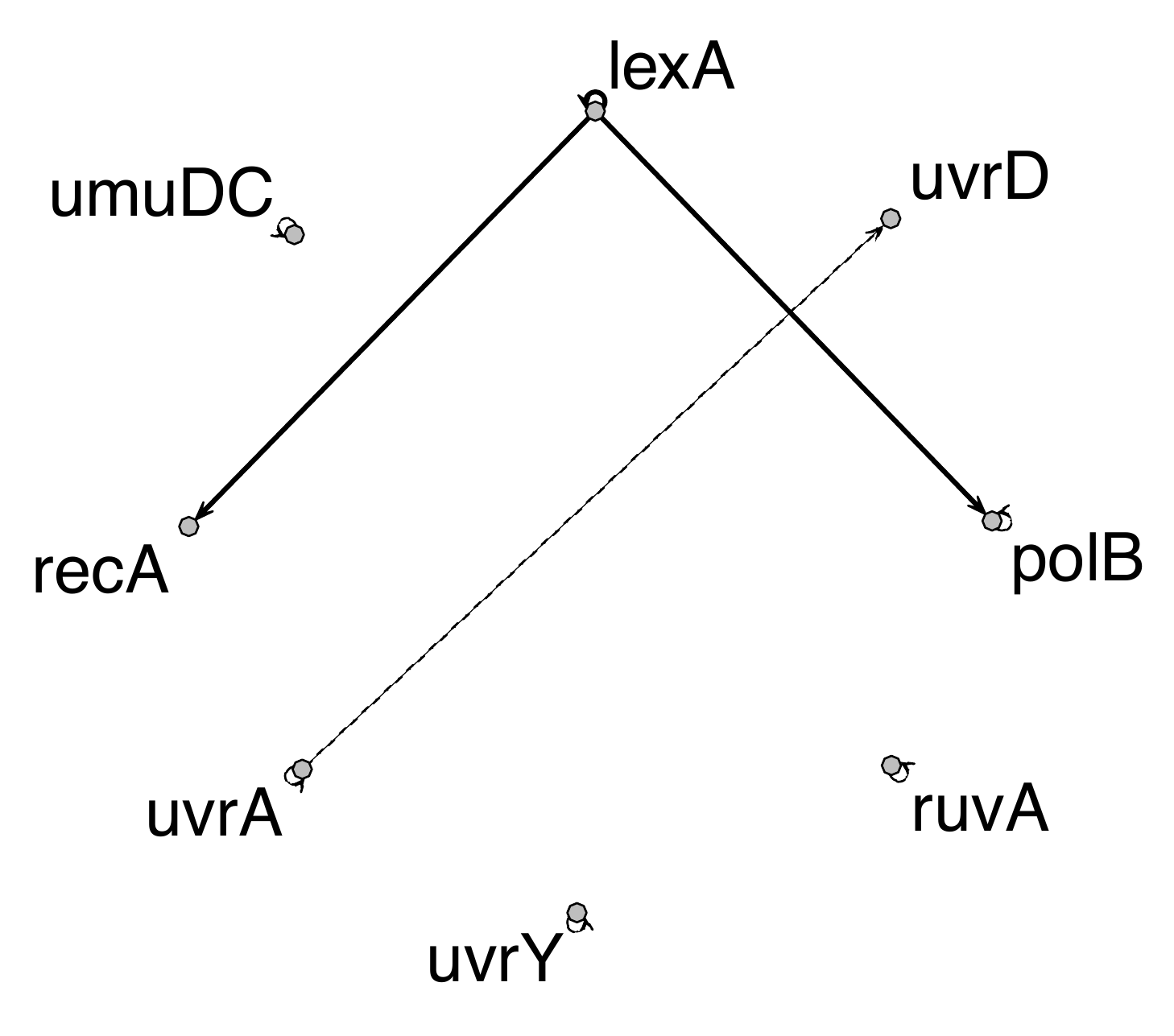}\\[3ex]

    \textsf{Known Classification} & \textsf{Inferred Classification}    \\ 
    \includegraphics[width=.4\textwidth]{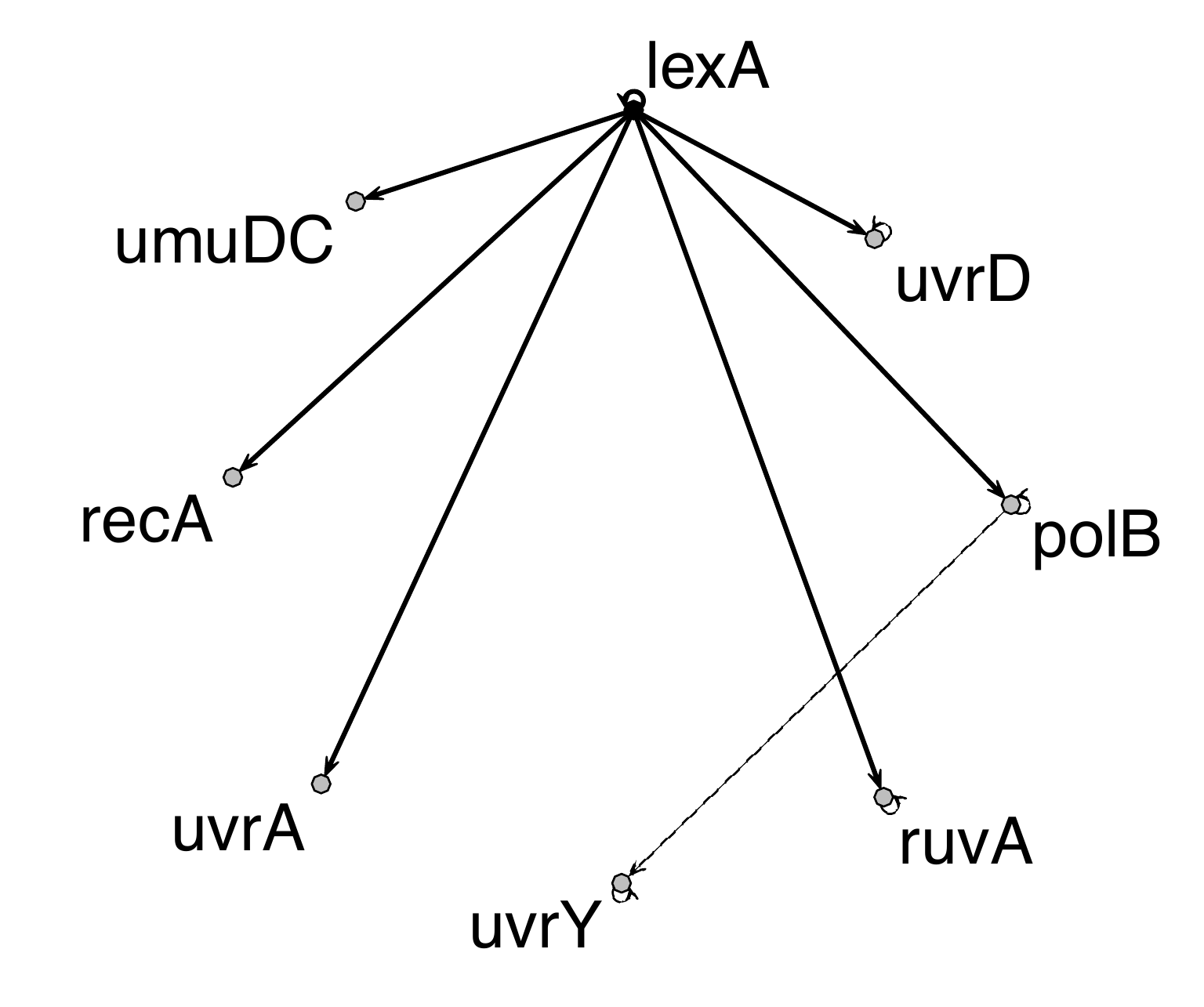} &
    \includegraphics[width=.4\textwidth]{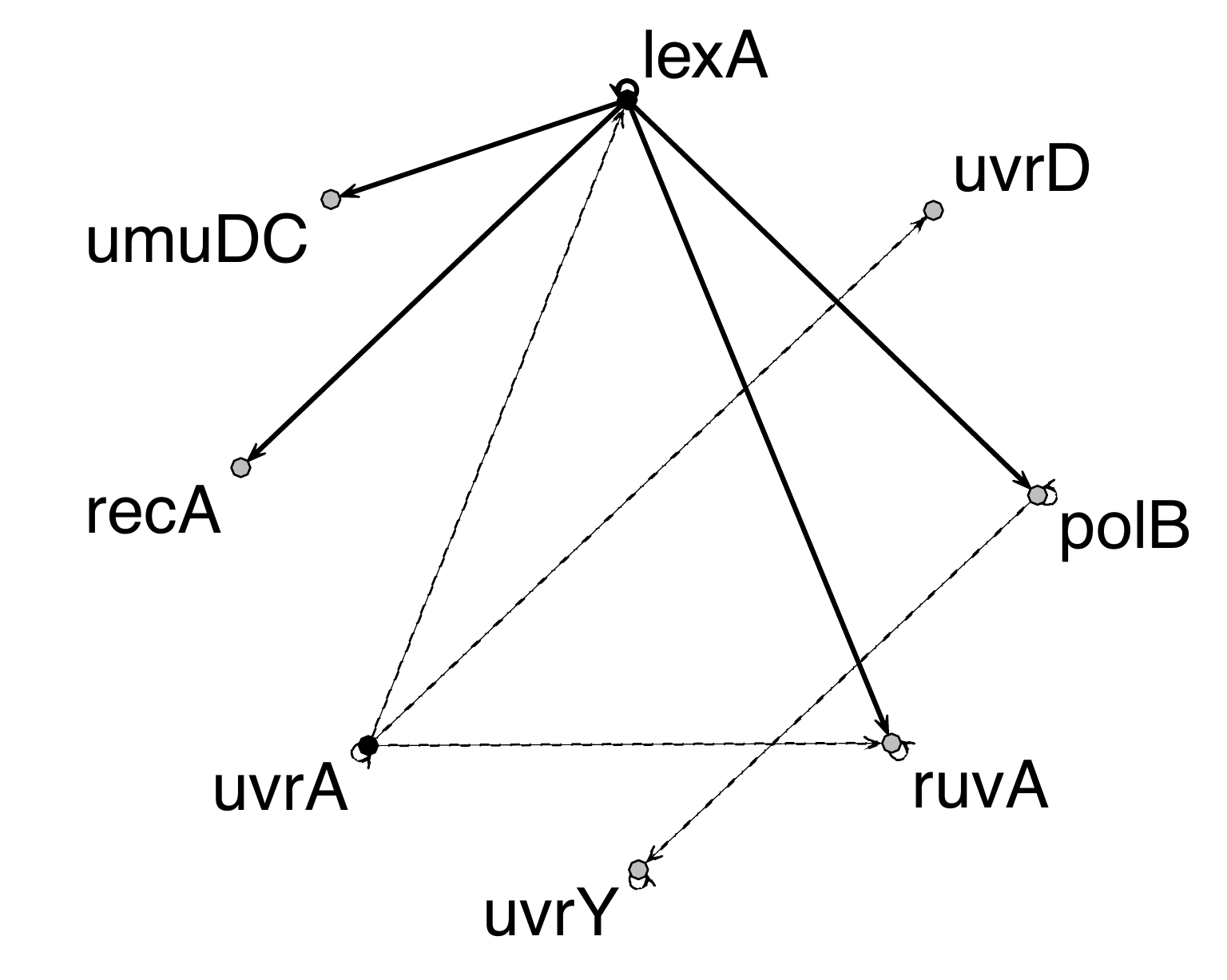}\\[3ex]

    \textsf{G1DBN} & \textsf{Renet-VAR} \\
    \includegraphics[width=.4\textwidth]{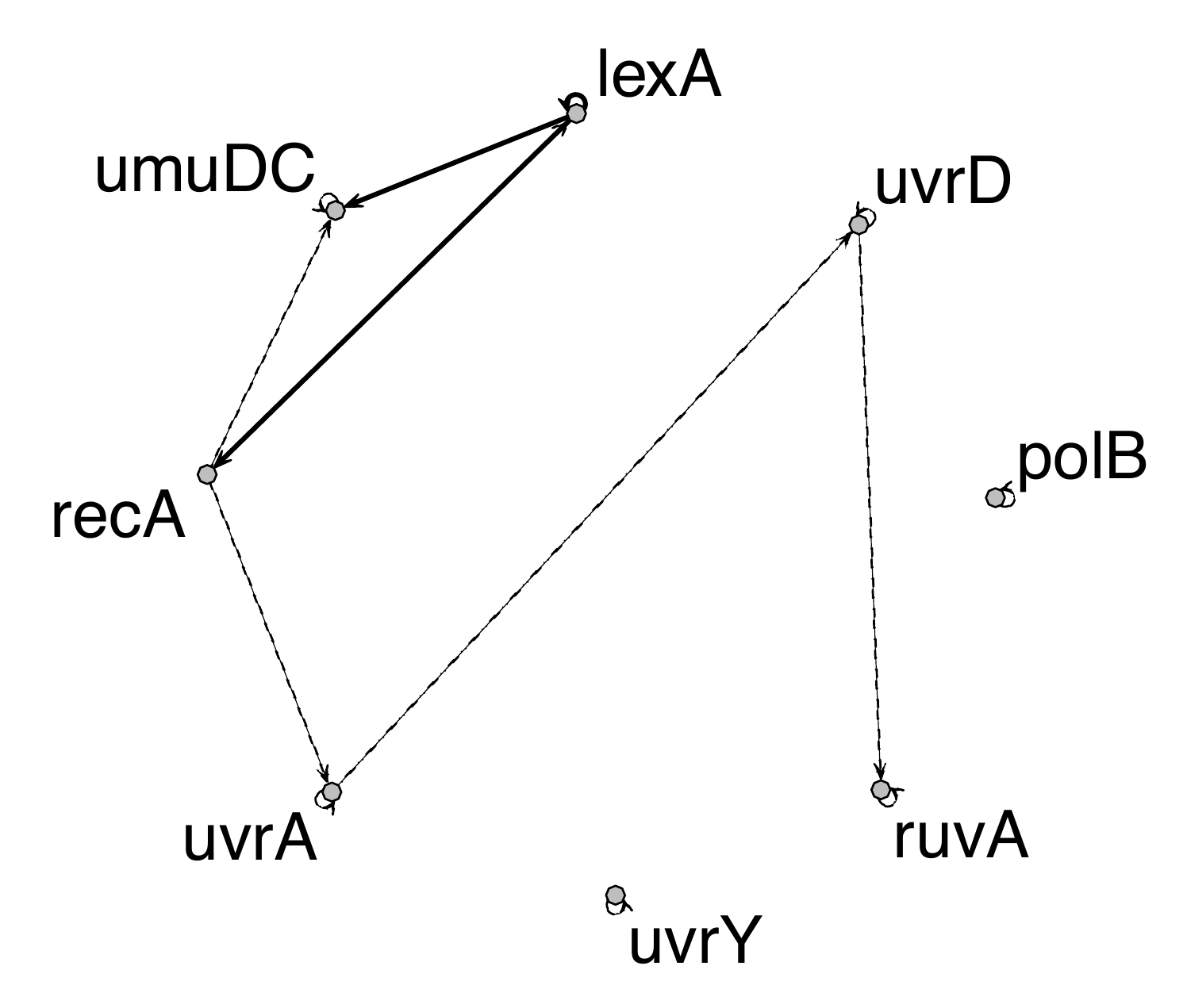} & \includegraphics[width=.4\textwidth]{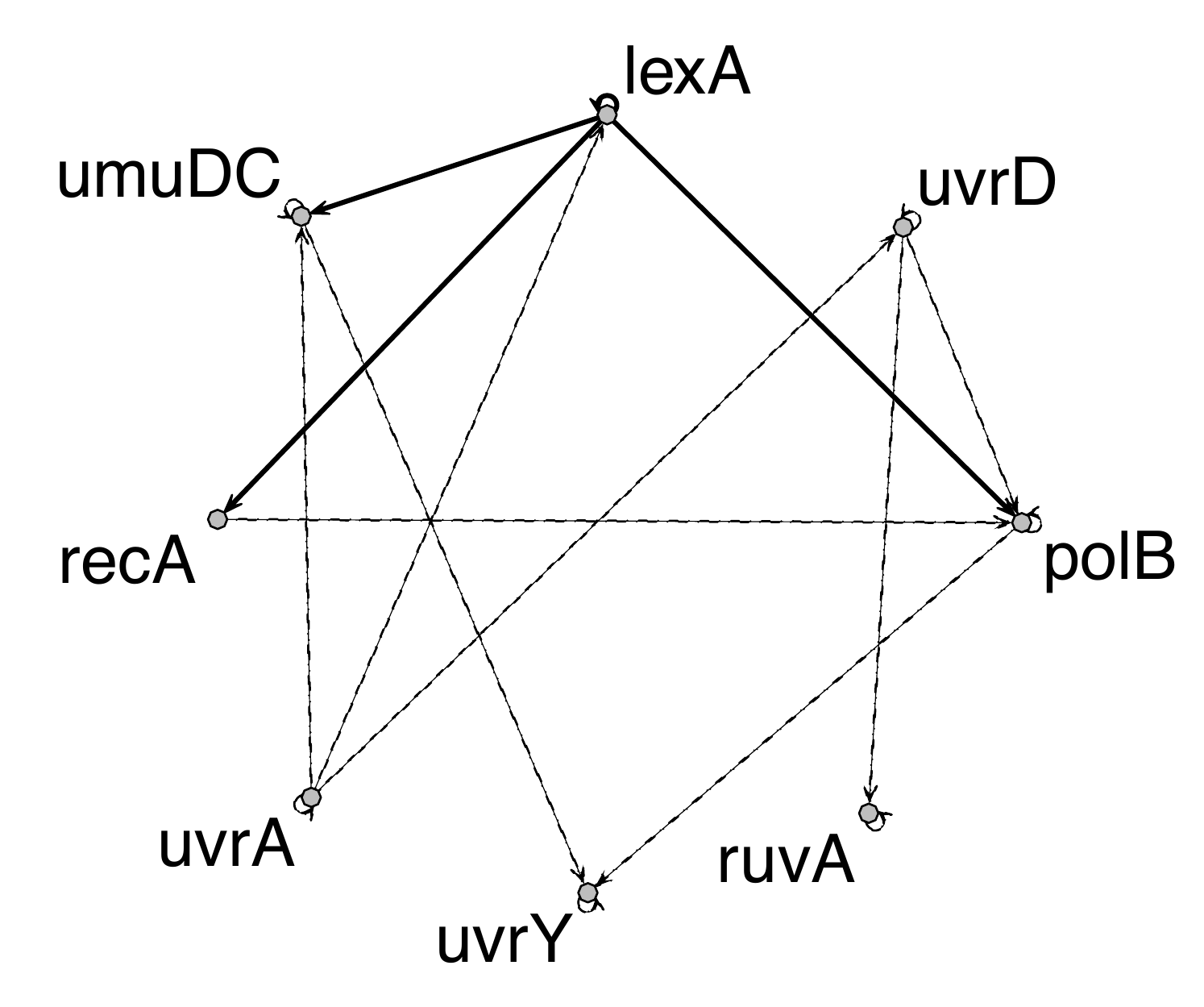}\\
  \end{tabular}
\end{small}
\caption{\label{fig:EcoliGraphs}  Graphs  inferred  by  the  different
  methods on experiment 2 data. \textsc{Lasso} penalties are chosen so
  as to maximize the BIC  criterion. True positives are drawn in black
  while false positives are shown in dashed gray.}
\end{figure}

Inferred graphs on experiment 2 are shown in Figure
\ref{fig:EcoliGraphs}. The regulatory activity of \emph{lexA} is more
or less recovered by all methods. What is interesting is that a common
structure recurently shows up among false positives: regulations
due to \emph{uvrA}. This regulation pattern is particularly what
dominates experiment 4 and leads to so poor results. Strangely, we
could not find any mention of this regulatory activity in the
literature. Either there is a need for further biological research on
this gene or there is an undirect regulation blurring the
results. Another unknown regulation dominates all inferred graphs:
regulation of \emph{uvrY} by \emph{polB}. It is all the more
interesting as it survives the bad \emph{a priori} that the
\emph{KnwCl} penalty holds against it. Further biological investigation
could want to look at this couple of genes more closely.

In this respect, we could note that the regulatory effect of
activated \emph{RecA} on \emph{LexA} does not appear on these graphs,
which we could see as a good point since this is a post-transcriptional
regulation. We would also like to lay the emphasis on the fact that we
here check selection consistency of all the methods but not their sign
consistency. We only check whether we identify the right edges and not
the activation/inhibition processes associated to them. Looking more closely at the estimated matrices, we can
see that the (shrunk) correlations estimated between \emph{lexA} and
the remaining genes are all positive and not negative as the
literature would tell. This would not be a flaw in all methods but a
direct result of the limitations of transcriptomic data. Indeed, we only observe mRNA production
rates. As a consequence, we cannot spot the decrease in concentration
of protein \emph{LexA} and only observe that the expression of all
genes suddenly increases, \emph{lexA} included. 

\section{Conclusion}

This  paper proposes a  weighted-\textsc{Lasso} algorithm  designed to
tackle  time  varying gene  expression  data  taking  into account  an
underlying  structure. We  observe  that in  a  perfect VAR1  setting,
taking  time  dependencies into  account  leads to dramatically  improved
results for graph inference.   In   this  particular  framework,   the  proposed  approach
outperforms similar methods. Even when regulators and regulatees
cannot \emph{a priori} been distinguished through analysis of the
literature, inference of the classification improves a lot the
performances of the \textsc{Lasso}. It therefore seems good to advice
that, whenever available, knowledge about potential transcription
factors should be taken into account and that basic knowledge on the
topology of biological networks should not be omitted in the modeling
process. We also want to emphasize the fact that this method reaches
great results on networks of reasonable size for always reasonnable computing times.


\bibliography{biblio}

\end{document}